STEFANO BETTINI (betstef@libero.it)

# A COSMIC ARCHIPELAGO: MULTIVERSE SCENARIOS IN THE HISTORY OF MODERN COSMOLOGY

ABSTRACT: Multiverse scenarios are common place in contemporary high energy physics and cosmology, although many consider them simply as bold speculations. In any case there is nothing like a single theory or a unified model of the multiverse. Instead, there are innumerable theoretical proposals often reciprocally incompatible. What is common to all these scenarios is the postulated existence of many causally disjoint regions of space/time (if not completely separated space-times) and the consideration of the observable universe as atypical in a global perspective. This paper examines the history of modern cosmology focusing on the forerunners of current ideas, and shows how some fundamental issues tend to present themselves on increasingly deeper levels of physical knowledge.

"Is it unavoidable to speak of these universes with different number of particles and (therefore!?) natural constants and h? That <u>must</u> be avoidable. If a theory presents the <u>possibility</u> of conceiving a theoretical deduction of the actual number of particles, i. e. of deducing, that N needs to be what it is, this theory <u>must</u> be capable of being described <u>without</u> using universes with an arbitrary number of particles. And I think, quite apart from this almost selfcontradictory procedure, it would greatly help the understanding to avoid the necessity of conceiving other universes than the actual one[1]".

"The idea that multiple domains may exist takes the Copernican revolution to its ultimate limit - even our universe may not be the center of the Universe[2]."

## 1. INTRODUCTION

The 63th Symposium of the International Astronomical Union, held in Cracow on September 1973, was largely dedicated to cosmological issues deriving from the general relativistic hot big bang (HBB) model of the universe[3]. It was a logical choice if one thinks that in that period, especially after the publication of textbooks by Peebles and Weinberg[4], the HBB had been canonized as the paradigmatic standard model of scientific cosmology.

On the other hand, it could appear a little bit surprising that at this same symposium at least three different participants referred to the possible existence of many universes distinct from our own.

Gary Steigman invoked "an ensemble of very many possible universes" in order to throw light on the excess of particles over antiparticles in our universe[5], while Hawking spoke of "conceivable universes"[6] while repeating the results of a recent paper (written jointly with Collins) dedicated to an explanation of the observed isotropy of the universe along the lines suggested by Robert Dicke and Brandon Carter[7].

---

[1] Schrödinger to Eddington, letter dated October 23th 1937, *Archives for the History of Quantum Physics - microfiche 37*.

[2] Agrawal, V.-Barr, S.M.-Donoghue, J.F.-Seckel, D.: *Anthropic Considerations in Multiple-Domain Theories and the Scale of Electroweak Symmetry Breaking*, Physical Review Letters 80, 1998, p. 1822/1825.

[3] Longair, M.S. (ed.): *Confrontation of cosmological theories with observational data*, Reidel, Dordrecht 1974. In the abstract of the Proceedings one can read that "the major issues studied" were: "structure and dynamics of the universe, the relic radiation, the origin of structure in the expanding universe, the structure of singularities, and the physical processes near a singularity … the choice between open and closed models, the non uniform distribution of matter (in particular, the clustering of galaxies), the origin of galaxies, and the rotation and magnetic fields of galaxies".

[4] Peebles, P.J.E.: *Physical Cosmology*, Princeton University Press, Princeton 1971; Weinberg, S.: *Gravitation and Cosmology*, Wiley, New York 1972.

[5] Steigman, G. 1974: *Confrontation of Antimatter Cosmologies with Observational Data*, in: *Confrontation of …*1974, p. 347/356 (quotation on p. 355).

The "Dicke/Carter philosophy"[8] was indeed presented in Cracow by Carter himself[9], who, upon the invitation of Wheeler, exposed a "line of thought" developed by him over the last seven years[10], finally introducing the term "anthropic" to re-name what was previously called the "principle of cognizability"[11].

In its strong form, the anthropic principle was then associated with an ensemble of universes and exploited to show how certain numerical coincidences, which depend critically "on the actual values of fundamental or microphysical constants"[12] (such as Sciama's relation between the average density and the square of Hubble constant at the present time[13] or the peculiar weakness of the gravitational constant), were indeed predictable without deserting conventional physics and cosmology. Carter considered the constants of nature as operators in an Everett/Hilbert space where "an ensemble of universes characterized by all possible combinations of initial conditions and fundamental constants" was represented[14].

To postulate the "existence" of such an ensemble could appear a bold move open to different kinds of objections, but according to Carter it did not bring "very much further than the Everett doctrine"[15] which represents "the only interpretation of quantum theory … which makes sense in cosmological contexts"[16].

In the late 1960's and early 1970's the three above mentioned contributions were accompanied by others of a similar kind or in any case discussing a "multiverse" concept[17]. Among these we find Bryce Seligman de Witt's presentation of Everett's relative state formulation of quantum mechanics in terms of a "continual splitting of the universe into a multitude of mutually unobservable but

---

[6] Hawking, S.W.: *The anisotropy of the universe at large times*, in: *Confrontation of …*, 1974, p. 283/286, on p. 285.

[7] Collins, C.B.-Hawking, S.W.*: Why is the universe isotropic?*, Astrophysical Journal 180, 1973, p. 317/334. The initial draft of the paper was received by the Journal on June 29, 1972. A revised version followed on September 25.

[8] Thus named by Collins and Hawking and by Wheeler in Misner, C.W.-Thorne, K.-Wheeler, J.A.: *Gravitation*, Freeman, San Francisco 1973, p. 1216.

[9] Carter, B.: *Large number coincidences and the anthropic principle in cosmology*, in: *Confrontation of …*, 1974, p. 291/298.

[10] Carter, B.: *The significance of numerical coincidences in nature. Part 1: the role of fundamental physical parameters in cosmogony*, DAMTP preprint, University of Cambridge 1967; Carter, B.: *Large numbers in astrophysics and cosmology*, unpublished preprint 1970. I have discussed the contents of these papers and the historical roots of the anthropic principle(s) in: Bettini, S.: *The Anthropic Reasoning in Cosmology: A Historical Perspective*, in: *Teleologie und Kausalität*, Stöltzner, M.-Weingartner, P. (eds.), Mentis, Paderborn 2005, p. 35/76 or arXiv:physics/0410144; Bettini, S.: *Il Labirinto Antropico*, http://www.swif.uniba.it/lei/saggi/antropico .

[11] Carter, B.: *Large number in astrophysics …*, 1970; Dyson, F.J.: *The fundamental constants and their time variation*, in: *Aspects of quantum theory*, Salam, A.-Wigner, E.P. (eds.), Cambridge 1972, p. 213/236. See also: Rees, M.J. 1972: *Cosmological significance of $e^2/Gm^2$ and related large numbers,* Comments on Astrophysics and Space Physics 4, 1972, p. 179/185; Harrison, E.R.: *Cosmological principles II. Physical principles*, Comments on Astrophysics and Space Physics 6, 1974, p. 29/35.

[12] Carter, B.: *Large number in astrophysics …*, 1970, p. 4 of the typescript.

[13] $G\rho_0 H_0^{-2} \approx 1$ See: Sciama, D.: *On the origin of inertia*, Monthly Notices of the Royal astronomical Society 113, 1953, p. 34/42.

[14] Carter, B.: *Large number in astrophysics …*, 1970, p. 5 of the typescript.

[15] Carter, B.: *Large number coincidences…*1974, on p. 298. Carter did not quote Everett's papers but: de Witt, B. S. 1967: *Quantum Theory of Gravity. I. The Canonical Theory*, Physical Review 160, 1967, p. 1113/1148.

[16] Carter, B.: *Large number in astrophysics …*, 1970, p. 10 of the typescript. Cf. Barrow, J.D.-Tipler, F.J: *The Anthropic Cosmological Principle.*, Oxford University Press, Oxford 1986, p. 496: "the wave function collapse postulated by the Copenhagen interpretation is dynamically ridiculous, and this interpretation is difficult if not impossible to apply in quantum cosmology".

[17] Popular essays on the multiverse were then available at the end of the decade. E.g. Davies, P.: *Other Worlds*, J.M. Dent, London 1980.

equally real worlds", which is perhaps the best known and the most controversial[18]; Wheeler's ideas on superspace[19] and about the cyclical universe[20]; Tryon's thesis concerning the creation of the universe out of nothing[21]; Thomas Gold's proposal of "other universes nesting within our observable space" in the form of closed subunits appearing "as very small areas of great redshift"[22]; Reinhard Breuer's and Michael Ryan's suggestion of "other metagalaxies" surrounding our own "within a larger universe"[23] and Tullio Regge's attempts to justify some dimensionless numbers invoking "other universes"[24].

Although there are relatively few physicists in the first half of the 1970's ready to speculate on many universes, amongst them we can find eminent names and scholars who almost certainly influenced the generations to come. Their pioneering works represented the prelude of a trend which assumed larger proportions in the early 1980's, when the elaboration of concepts such as those of phase transitions and spontaneous symmetry breaking threw new light on the role of intrinsically random processes in the very early universe[25]. The possibility that fundamental properties of the universe, including the dimensionality of space/time and the values of the fundamental constants, might have a probabilistic origin was then seriously considered, opening the doors to speculations about other regions, bubbles, domains, branches, pocket-universes or simply "universes".

Cosmologists were since then motivated to theorize beyond the standard model for reasons which span from private facts (such as a "feeling of excitement"[26] or the need to publish in an ever expanding global flourishing of new publications and new media) to undoubtedly more cogent motives. Among the latter we should maybe include:
- the need to solve the open problems of the HBB model (beginning with the problems of flatness and horizon, which were already known in the late 1960's[27], but which were largely appreciated after the 1979 paper by Dicke and Peebles [28] and which then generated Guth's inflationary model[29])
- the need to explore the physics of the very early universe, going beyond the standard models of both cosmology and elementary particles (and developing new mathematical techniques while facing the extreme conditions between the hadron era and Planck time)

---

[18] De Witt, B.S.-Graham, S. (eds): *The Many-Worlds Interpretation of Quantum Mechanics*, Princeton University Press, Princeton, 1973 (quotation on p. 5); Cf. also: de Witt, B.S.: *Quantum Mechanics and Reality*, Physics Today 23, No. 9 (September 1970), p. 30/40 and the debate which followed in: Physics Today 24, No. 4 (April, 1971).
[19] E.g.: Wheeler, J.A: *Our universe: the known and the unknown*, American Scientist 56, 1968, p. 1/20.
[20] E.g.: Misner, C.W.-Thorne, K.-Wheeler, J.A.: *Gravitation*, …, 1973, section 44.6; Patton, L.M.–Wheeler, J.A.: *Is physics legislated by cosmogony?*, in: *Quantum gravity: an Oxford symposium*, Isham, C.J.-Penrose, R.-Sciama, D. W. (eds.), Oxford 1975, p. 538/605; Wheeler, J.A: *Genesis and observership*, in: *Foundational problems in the special sciences*, Butts, R.E.-Hintikka, J. (eds.), Dordrecht 1977, p. 3/33. This topic will be reconsidered in part 4 below.
[21] Tryon, E.P.: *Is the universe a vacuum fluctuation?*, Nature 246, 1973, p. 396/397
[22] Gold, T. 1973: *Multiple Universes*, Nature 242, p. 24/25
[23] Breuer, R.A.-Ryan, M.P.: *The porthole effect and rings of fire in finite metagalaxy*, Monthly Notices of the Royal Astronomical Society 171, 1975, p. 209/218. Breuer was one of Sciama's student and the author of an early popular essay on the anthropic principle. See: Breuer, R.: *Das antrhopische Prinzip: der Mensch im Fadenkreuz der Naturgesetze*, Meyster, Wien/München 1981.
[24] Regge, T.: *S-matrix Theory of Elementary Particles*, in: *Atti del convegno Mendeleeviano. Periodicità e simmetrie nella struttura elementare della materia. Torino-Roma 15-21 settembre 1969*, Verde, M. (ed.), Accademia delle Scienze di Torino, 1971, p. 395/401 (quotation on p. 398).
[25] Cf.: Barrow, J.: *The World within the World*, Clarendon Press, Oxford, 1988; *Unprincipled Cosmology*, Quarterly Journal of the Royal astronomical Society 34, 1993, p. 117/134.
[26] Cf.: Moss, I. G.: *Quantum Theory, Black Holes and Inflation*, Wiley, New York 1996, p. 32
[27] E.g. Dicke, R.H.: *Gravitation and the Universe. The Jayne Lectures for 1969*, American Philosophical Society, Philadelphia 1970 (in particular p. 62).
[28] Dicke, R.H.-Peebles, J.: *The Big Bang Cosmology - Enigmas and Nostrums* in: *General Relativity: an Einstein Centenary Survey*, Hawking, S.W.-Israel, W. (eds.), Cambridge University Press, Cambridge 1979, p. 504/517.
[29] Guth, A.H.: *Inflationary Universe: A Possible Solution to the Horizon and Flatness Problems*, Physical Review D23, 1981, p. 347/356.

- the need to point to several broader scenarios endorsing the principle of plenitude[30] (at the same time confronting ideas, principles or beliefs which have a long history behind them)
- the need to solve the problem of fine tuning

The apparent "fine tuning" of many physical and cosmological quantities in the observed region of the universe represents indeed a major conundrum. It seems, in fact, that the existence of all kinds of complexity (including, of course, intelligent life) fundamentally depends on the "delicate balance" between the values of several physical and cosmological quantities (from parameters of the standard model and fundamental physical constants[31] to the dramatic case of the cosmological constant, which is roughly 120 orders of magnitude smaller than the value that particle theorists would expect[32]). This puzzling situation constitutes one of the main motivations for the connection between multiverse scenarios and "anthropic reasoning"[33]. In short, it does not matter how improbable the values of physical and cosmological fundamental quantities are *a priori*, since we must necessarily live in one of the universes that allows life.

Both anthropic reasoning and many universes have become common occurrence nowadays. But this means that, as in the case of the anthropic principle(s)[34], the appeal to the multiverse concept implies a rather chaotic state of affairs, which includes different and often incompatible scenarios. Paraphrasing John Earman[35], one could say that the multiverse is not a single and unified conception but rather a complicated network of theoretical proposals or speculations which are often in reciprocal antagonism.

As a matter of fact, in the current literature one can find very different kinds of multiverses and innumerable mechanisms to generate them (as for instance: oscillations, quantum world-splitting or other forms of foliation of Hilbert space, quantum vacuum fluctuations, symmetry breaking bubbles production, chaotic distributed scalar fields, inflaton's fields, wormholes, Smolin's black holes, random difference of chaotic gauge theories, moduli space of solutions of a supersymmetric M-theory, primordial antimatter domains at high redshifts[36], ... not to speak of the idea of producing universes in the laboratory through "a small region of false vacuum"[37], of cosmic organism's theories[38] or of the suggestion concerning fake " simulated universes" indistinguishable from real ones in the eyes of their virtual inhabitants[39]). Moreover universes can generally differ not only in size or free parameters typical of the HBB model, but also in strength of elementary forces, masses of elementary particles, gauge symmetries, vacuum energy densities (i.e.: values of the cosmological constant), metric signatures, dimensionality and so on, thus frustrating almost any attempt of looking for an accurate taxonomy of the existing multiverse scenarios[40].

---

[30] According to Arthur O. Lovejoy (who does not forget God's Creative omnipotence) the theorem of plenitude states that all the conceptual possibilities must be realized. Cf. Lovejoy, A.: *The Great Chain Of Being: A Study of the History of an Idea*, Harvard University Press, Cambridge, Mass. 1936.

[31] For a huge list see: Ross, H.: *The Creator and the Cosmos: How the Greatest Scientific Discoveries of the Century Reveal God*, second edition., Colorado Springs, Navpress, 1995, p 118/21 or Davies, P.: *The Accidental Universe*, Cambridge University Press, Cambridge 1982.

[32] Among the available reviews: Carroll, S,-Press, W.-Turner, E.: *The Cosmological Constant*, Annual Review of Astronomy and Astrophysics 30, 1992, p. 499/542; Garriga, J.-Vilenkin, A.: *Solutions to the cosmological constant problems*, Physical Review D64, 2001, 023517 or arXiv:hep/th/0011262; Sahni, V.-Starobinsky, A.: *The Case for a Positive Cosmological Lambda-term*, International Journal of Modern Physics D9, 2000, p. 373/444; Weinberg, S.: *The Cosmological Constant Problems*, arXiv:astro-ph/0005265 (May 2000); Padmanabhan, T: 2003: *Cosmological Constant- The Weight of the Vacuum*, Physics Reports 380, p. 235/320. Early papers looking for an anthropic solution of the cosmological constant problem are: Davies, P.A.M.-Unwin, S.D.: *Why is the cosmological constant so small?*, Proceedings of the Royal Society A377, 1981, p. 147/149; Hawking, S.W.: *The cosmological constant and the weak anthropic principle*, in: *Quantum structure of space and time*, M. J. Duff-Isham, C. J. (eds.), Cambridge University Press, Cambridge 1982, p. 423/432; Hawking, S.W.: *The Cosmological Constant*, Philosophical Transactions of the Royal SocietyA310, 1983, p. 303/310; Weinberg, S.: *Anthropic Bound on the Cosmological Constant*, Physical Review Letters 59, 1987, p. 2607/2610.

[33] A common expression after Rees, M.J.: *Before the beginning. Our universe and others*, Simon & Schuster, London 1997.

[34] On the plural "principles" and on the controversial status of the different "forms" of the anthropic principle, see: Bettini, S.: *The Many Faces of the Anthropic Principle*, in: *Cosmology Through Time*, Colafrancesco, S.-Giobbi, G. (eds.), Mimesis, Milano 2003, p. 271/282.

[35] Earman wrote that the anthropic principle "is not a single, unified principle but rather a complicated network of postulates, techniques, and attitudes" (Earman, J.: *The SAP also rises: a critical examination of the anthropic principle*, American Philosophical Quarterly 24, 1987, p. 307-317, on p. 307). Cf. John Leslie's statement according to which: "there is just no single correct way of counting universes and thus of distinguishing them from mere regions or locations" (Leslie, J.: *Universes*, Routhledge, London 1989, p. 135).

Apart from all this, it seems that what is common to all the many universes consists in two essential elements: a causal separation between the universes and a probabilistic basis to establish what is typical or atypical to the whole ensemble. Both such elements appear relatively clear until one decides to scrutinize them more closely.

For instance, what is intended when mentioning causally disjointed universes? Probably regions of space/time which are beyond the horizon now (admitted that "now" could have a meaning outside of the hypersurface of simultaneity of an homogeneous universe)? Or, rather, domains that will remain forever disconnected? Or separated space/times that do not have a common causal origin in the past?[41] It is clear that many of the regions pictured in proposals which are commonly presented as multiverse scenarios are indeed parts of a unique space/time and thus locally causally related; nevertheless, we have reason to affirm that if regions of space/time are unobservable and beyond testability, they can then be considered as effectively disjointed (especially if one supposes that they present different physical and/or cosmological properties with regard to our observable region).

---

[36] Cf.: Dolgov, A.D: *Antimatter in the universe*, arXiv:hep-ph/9211241; *Matter-antimatter domains in the universe*, Nuclear Physics B - Proceedings Supplements 95, 2001, p. 42/46.

[37] Farhi, E.-Guth, A.H.: *An obstacle to create a universe in the laboratory*, Physics Letters B183, 1987, p. 149/155; Blau S. K.-Guendelman, E.I.-Guth A.: *Dynamics of false vacuum bubbles*, Physical Review D 35, 1987, p. 1746/1766; Farhi E.-Guth A.H.-Guven, J.: *Is it possible to create a universe in the laboratory by quantum tunneling?*, Nuclear Physics B339, 1990, p. 417/490; Guth, A.H.: *Creation of a Universe in the Laboratory*, in: *Relativity and Gravitation: Classical and Quantum, Proceedings of the 7th Latin American Symposium on Relativity and Gravitation,* D'Olivo, J.C. Nahmad Achar, E.-Rosenbaum, M.-Ryan Jr, M.P.-Urrutia, L.F.-Zertuche, F. (eds.): *World Scientific, Singapore*, 1991, p. 603/622; Linde, A.: *Hard Art of the Universe Creation*, Nucl. Phys. B372, 1992, p. 421/442; Guth, A.H.: *The inflationary universe: the quest for a new theory of cosmic origins*, Addison-Wesley, Reading (Massachusetts), 1997, chapter XVI. Probably also: Stapledon, O. 1937: *Star Maker*, Methuen, London.

[38] Chung, D.Y.: *The Cosmic Organism Theory for the Multiverse*, arXiv:hep-th/0201115.

[39] Bostrom, N.: *Are you living in a computer simulation?*, Philosophical Quarterly 57(211), 2003, p. 243/255 or http://www.simulation-argument.com ; Hanson, R.: *How to Live in a Simulation*, Journal of Evolution and Technology 7, 2001, in: http://hanson.gmu.edu/lifeinsim.html ; Davies, P.C.W.: *Multiverse Cosmological Models*, Modern Physics Letters A19, 2004, p. 727/744; Barrow, J.D.: *Living in a Simulated Universe, 2003, in: http://www.simulation-argument.com/barrowsim.pdf* . These works were anticipated by papers presenting the analogy Universe/computer. E.g.: Zuse, K.: *Rechnender Raum*, 1969 (now available in http://www.idsia.ch/~juergen/digitalphysics.html ); Wheeler, J.A.: *The Computer and the Universe*, International Journal of Theoretical Physics 21 (6/7), 1982, p. 557/572.

[40] E.g.: Barrow, J.D.-Tipler, F.J: *The Anthropic Cosmological ...*, 1986; Gale, G.: *Cosmological Fecundity: Theories of Multiple Universes*, in: *Physical Cosmology and Philosophy*, Leslie, J. (ed.) Macmillan, New York 1990, p. 189/206; Leslie, J.: *Universes ...* 1989; Leslie, J.: *Cosmology-A Philosophical Survey*, Philosophia: Philosophical Quarterly of Israel 24, 1993, p. 3/23; Vaas, R.: *Time before Time-Classifications of universes in contemporary cosmology, and how to avoid the antinomy of the beginning and eternity of the world*, Bild der Wissenschaft 10, 2004, p. 32/41 or arXiv:physics/0408111**.** Cf. also: Macleod, A.: *Coexisting Spacetimes in the Solar Neighborhood*, arXiv:physics/0506040; Bjorken, J.D.: *The classification of universes*, arXiv:astro-ph/0404233; Smolin, L.: *Scientific alternatives to the anthropic principle*, arXiv:hep-th/0407213. Contemporary classifications are in debt with the medieval one which (after Etienne Temper's decree of 1277) distinguished between three classes of universes: -a succession of single worlds in time -the simultaneous co-existence of worlds within other worlds -the co-existence of separated (and physically different) worlds within an infinite imaginary void space. The classic source on Medieval cosmology remains: Duhem, P.: *Medieval Cosmology: Theories of Infinity, Place, Time, Void, and the Plurality of Worlds*, University of Chicago Press, Chicago 1985.

[41] These questions resemble Hermann Bondi's distinction of three possible meanings of the term "universe": a) the totality of all objects to which physical theory can be applied whether these objects are observable or not; b) all objects that are observable now or at any time in the future; c) all objects that are now observable. Cf. Bondi, H.: *Some philosophical problems in comology*, in: *British philosophy in the mid century*, Mace, A. (ed.), George Allen and Unwin, London 1963, p. 393/400, on p. 394.

It is then very hard to say what makes a member of an ensemble of universes typical. What we generally do is to postulate the existence of typical outcomes in order to conclude that our observable region is atypical (although it remains possibly typical within an anthropic subset[42]).

The point is that one has no sure criteria to state what is common to every member of an ensemble, not even if one limits oneself to an anthropic subset. How do we delimitate such an ensemble? One can only presume that we have exhausted all the possibilities without being absolutely certain that this is the case.

In other words, a big problem of multiverse scenarios consists in defining a measure of probability over the members of an ensemble of universes, i.e. a measure indispensable to calculate probabilities concerning the parameters under scrutiny[43].

In practice, technical papers have limited themselves to probability distributions of one (or at most two[44]) parameters, posing convenient assumptions while nothing is said of more general cases.

Anyway, the present paper does not aim to discuss epistemological issues neither is it interested in analysing contemporary multiverse proposals. What it, however, intends to show are the historical antecedents of current scenarios.

## 2.MANY UNIVERSES IN PRE-RELATIVISTIC PHYSICS AND ASTRONOMY

The expression "island universes" appeared frequently in astronomical debates from the times of Lord Rosse to the years of the Great Debate between Curtis and Shapley. At any rate, this important chapter in the history of astronomy is of meagre relevance to our current discussion. The "island universes" dispute, in fact, essentially dealt with the nature and collocation of those spiral nebulae which were observed copiously since William Herschel[45]. The point, as everyone knows, was to establish if objects of such a kind were remote islands of stars comparable in size to our Milky Way or if, on the contrary, they were mere subunits of our own Galaxy. In the latter case, which was considered orthodox at the turn of XIX century, the Milky Way was generally seen as a finite aggregate of stars immersed in an infinite Euclidean space and this fundamental "Architecture of the Heavens"[46] was treated as the Universe in its own right[47].

Nevertheless, the debate on the nature of the Nebulae stimulated a long series of speculations about very remote and unobservable outer universes. Whatever the nature of the observed Nebulae, imagining the existence of other universes of stars which are very far away (possibly so far away to have no influence on our neighbourhoods) was indeed an almost inevitable consequence of the principle of plenitude.

---

[42] The task of "adequately specifying possible anthropic universes" has been attempted in: Cf. Ellis, G.F.R.–Kirchner, U.–Stoeger, W.R.: *Multiverses and Physical Cosmology*, Monthly Notices of the Royal Astronomical Society 347, 2004, p. 921/936 or arXiv:astro-ph/0305292. See also: Stoeger, W.R.-Ellis, G.F.R.–Kirchner, U.: *Multiverses and Cosmology: Philosophical Issues*, arXiv:astro-ph/0407329 (July 2004).

[43] Cf. Aguirre, A.: *On making predictions in a multiverse: conundrums, dangers, and coincidences*, arXiv:astro-ph/0506519 (June 2005); Aguirre, A.-Tegmark, M.: *Multiple universes, cosmic coincidences, and other dark matters*, Journal of Cosmology and Astroparticle Physics 0501, 2005, 003, or: arXiv:hep-th/0409072 (September 2004).

[44] E.g.: Garriga, J.–Vilenkin, A.: *Anthropic prediction for Λ and the Q catastrophe*, arXiv:hep-th/0508005 (August 2005).

[45] Cf: Crowe, M.J.: *Modern Theories of the Universe. From Herschel to Hubble*, Dover, New York 1994; Dewhirst, D.W.-Hoskin, M.A.: *The Rosse Spirals*, Journal for the History of Astronomy 22, 1991, p. 257/266; Hoskin, M.A.: *Apparatus and Ideas in Mid-nineteenth-century Cosmology,* Vistas in Astronomy 9, 1967, p. 79/85; Hoskin, M.A.: *Stellar Astronomy: Historical Studies*, Science History Publications, Chalfont St. Giles 1982; Hoskin, M.A.: *Rosse, Robinson, and the resolution of the nebulae*, Journal for the History of Astronomy 21, 1990, p. 331/344; Jaki, S.L.: *The Milky Way. An Elusive Road for Science*, Science History Publications, New York 1973; Merleau-Ponty, J.: *La science de l'univers a l'âge du positivisme. Etude sur les origines de la cosmologie contemporaine*, Vrin, Paris 1983.

[46] Nichol, J.P.: *Views of The Architecture of the Heavens in a Series of Letters to a Lady*, William Tait, Edinburgh 1837.

[47] This view was canonized in books such as: Clerke, A.: *The System of the Stars*, Longmans, Green & Co., London 1890 or Newcomb, S.: *Side-Lights on Astronomy. Essays and Addresses*, Harper and Brothers, New York 1906, now available on: http://www.gutenberg.org/dirs/etext03/slstr10.txt .

In an epoch of strong positivism, when observability was synonymous with scientific existence, these bold speculations were generally relegated outside the circles of professional astronomers. It is not, therefore, surprising if very early suggestions of multiverse scenarios are the products of men of letters or amateurs who were free of speculating beyond the "observed facts".

Think, for instance, of Poe's lectures on cosmology which were printed in 1848. The famous essay, *Eureka-A Prose Poem*[48], was presented by its author as a "book of truths" dedicated to dreamers and contained numerous insights which appear almost prophetical in retrospect.

Once the description of our universe as a "cluster of clusters" governed by unitary laws and subjected to a peculiar evolutionary history had been introduced, the American writer depicted "an interminable succession of "clusters of clusters", or of "Universes" more or less similar". He declared that such a possibility was just a "fancy", but at the same time stated that:

"If such clusters of clusters exist, however -- and they do -- it is abundantly clear that, having had no part in our origin, they have no portion in our laws. They neither attract us, nor we them. Their material -- their spirit is not ours -- is not that which obtains in any part of our Universe. They could not impress our senses or our souls. Among them and us -- considering all, for the moment, collectively -- there are no influences in common. Each exists, apart and independently, in the bosom of its proper and particular God".

Contemporary scholars have seen an anticipation of the idea of causally disjointed universes in these lines, recognizing in Poe's words a clear suggestion of the existence of other cosmoi rather than of other galaxies[49].

A fashionable book such as *Eureka* was probably perceived as a product of imagination by astronomers of the second half of XIX century, but the absence of a causal connection was a theme that occasionally arose within some of the many debates concerning the nature of aether.

In a popular exposition, the *Lezioni di Astronomia*, published in 1877 by Giuseppe Barilli (who was in the habit of signing his books under the name of Quirico Filopanti) one can find mention of the existence of "innumerable Cosmoi, physically segregated from each other and from ours". Filopanti was almost certain of their existence, although it was impossible to detect them because of "the lack of an interposed aether" or - in other words – of a suitable medium apt to the transmission of light[50]. The Italian "popular lecturer" didn't exclude that some fragments of such remote Cosmoi could have occasionally travelled through the depth of space to our cosmic neighboorhood under the form of bolides, meteorites or other wandering astronomical bodies (a topic which assumed a proper relevance in astronomical debates after the discovery of the unusually high proper velocities of some "roving stars"[51]). He argued moreover that our position within the Milky Way (thus surrounded by other stars and nebulae in every direction) favoured our existence with regard to the possible occurrence of cosmical catastrophes associated with colliding astronomical debris.

---

[48] Poe, E.A.: *Eureka. A Prose Poem*, Putnam, New York 1848, now available on http://xroads.virginia.edu/~HYPER/poe/eureka.html .

[49] Van Slooten, E.R.: *Poe's universes*, Nature 323, 1986, p. 198; Tipler, F.J.: *Johann Mädler's Resolution of Olber's Paradox*, Quarterly Journal of the Royal Astronomical 29, 1988, p. 313/325; Cappi, A.: *Edgar Allan Poe's Physical Cosmology*, Quarterly Journal of the Royal Astronomical Society 35, 1994, p. 177/192.

[50] Filopanti, Q.: *Lezioni di Astronomia*, L. Bortolotti, Milano 1877 (quotation on p. 30). Filopanti stated that "no one would dare deny" the existence of such an "obscure desert" populated by other Cosmoi beyond our own. His contemporary, Padre Angelo Secchi, rejected anyway such an absolute vacuum inviting his readers to "dream not of what we can not see". Cf.: Secchi, A.: *Le stelle. Saggio di astronomia siderale*, Fratelli Dumolard, Milano 1878, p. 336/337.

[51] Newcomb, S.: *Popular Astronomy*, Macmillan & Co., London 1878 (cf. In part. p. 487/488); Thomson, W. (Lord Kelvin): *On Ether and Gravitational Matter through Infinite Space*, Philosophical Magazine (6) 2, 1901, p. 161/177. An "etherodox" explanation of these runaway stars, involving negative mass, was given in: Pearson, K.: *Ether-squirts; Being an Attempt to Specialize the Form of Ether-Motion which forms an Atom in a Theory propounded in former Papers*, American Journal of Mathematics 13, 1891, p. 309/362.

Finally, Filopanti invited his contemporaries to use the term "cosmos" as synonymous of the observable universe, while reserving the word "Universe" for that "supreme" infinite entity which includes all the innumerable Cosmoi.

The idea of a cosmic "desert" devoid of ether and a scenario very similar to that described in the *Lezioni d'Astronomia* was presented some years later by the Irish astronomer John Ellard Gore and exploited as a possible resolution of Olber's paradox. Still in the 1904 edition of his *Studies in Astronomy*, Gore found, indeed, reasons to describe our star system as

"a limited universe, which is isolated by a dark and starless void from any other universes which may exist in the infinity of space beyond"[52].

Although Filopanti's and Gore's speculations can appear today as marginal contributions in a historical perspective, the debates which arose after the formulation of the second law of thermodynamics represent a different case. In the first place because this law (or theorem, which is a more precise translation of the German word *Hauptsatz*) included the whole universe since the very beginning. As Clausius put it in 1868:

"The more the universe approaches this limiting condition in which the entropy is a maximum, the more do the occasions of further changes diminish; and supposing this condition to be at last completely obtained, no further change could evermore take place, and the universe would be in a state of unchanging death".[53]

A big question left open by such a formulation of the second *Hauptsatz* concerned the actual state of the observed world[54]. Given enough time (or better still the eternity of time that many authors were inclined to assume), why do our surroundings testify to a state of things which is very far from the presumed Heath Death? A similar question was, for instance, explicitly posed by Johannes Gustav Vogt in the first book of a treatise entitled *Die Kraft*. Vogt asked:

"Since the world is temporally infinite, why has this limiting condition [of maximum entropy] not been reached a long time ago?"[55]

If one wishes to avoid recourse to very special initial conditions or the disturbing idea of a beginning of the whole world which seems to blatantly contradict the first of the laws of thermodynamics, it is necessary to individuate other ways out of this problem. A distinctive solution was advanced by an English expert in kinetic theory responding to the name of Samuel Tolver Preston[56] who, in 1879, compared human observers to a Maxwell's demon intent on reaching conclusions about the equilibrium temperature of a gas from the point of view of a single molecule[57]. Preston argued that the scale of the universe was simply "too big" for our standards and suggested that the region of the universe that we inhabit, with its „rather exceptional" concentration of hot luminous stars, was presumably atypical if seen in a truly global perspective. He suggested then that the presence of regions where „the conditions necessary for life" are maintained for periods of time which are long in comparison to the human experience are consistent with the second law, recalling that the fact of finding ourselves in a part of the universe "where the mean temperature of the component matter is exceptionally high" reserves no surprise if one considers that "from the fact of our being in existence, we must be in a part which is suited to the conditions of life"[58].

---

[52] Gore, J.E. 1904: *Studies in Astronomy*, Chatto and Windus, London 1904, p. 137.

[53] Clausius, R.: *On the Second Fundamental Theorem of the Mechanical Theory of Heat; a Lecture delivered before the Forty-first Meeting of the German Scientific Association, at Frankfort on the Maine, September 23, 1867*, Philosophical Magazine (4) 35, 1868, p. 405/419 (on p. 419).

54 Clausius stressed that "the present condition of the universe" was "still very far removed" from the limiting condition. He then remarked that "all such periods as we speak of as historical are but a very short span in comparison with the immeasurable periods that the universe requires for comparatively very slight modifications".

[55] Vogt, J.G.: *Die Kraft. Eine real monistische Weltanschauung. Erstes Buch. Die Contraktionsenergie, die letztursächliche einheitliche mechanische Wirkungsform des Weltsubstrates*, Hautp & Tischler, Leipzig 1878. An obvious alternative consists in a universe of finite age starting in a very special state, but creation and special initial conditions were unacceptable for Vogt as they will be for Boltzmann's mechanical *Weltbild*. Opposite views are discussed in: Landsberg, P.T.: *From Entropy to God?* in: *Thermodynamics: History and Philosophy*, Martinas, K.-Ropolyi, L.-Szegedi, P. (eds.), World Scientific, Singapore 1991, p. 379/403; Kragh, H.: *Matter and Spirit in the Universe, Scientific and Religious Preludes to Modern Cosmology*, Imperial College Press, London 2004.

[56] Samuel (b. 1844) was the son of Daniel Bloom Preston and Mary Susannah Tolver. Cf. http://stanfield.und.ac.za/gedhtree/oldjermy/gp2260.html

Preston's explanation was accused of arousing „confusion of reasoning" and "unsoundness", but the author defended his perspective affirming that it was "consistent with principles which at present prevail" and escaped succesfully from the "necessity of supposing that existing physical principles must have been violated in past time" [59].

The problem of the actual state of the world continued anyway to repropose itself. Nothwithstanding the enormous developments that characterized the fields of kinetic theory and statistical mechanics in the second half of the XIX century (especially thanks to the powerful mathematical techniques developed by Maxwell and Boltzmann), in 1894 Fitzgerald was still there to ask:

"… why the ether, the solar system, and the whole universe were not subject to the Boltzmann/Maxwell law?"[60]

As is well known, it was Boltzmann himself who provided an answer which fundamentally echoed Tolver Preston's argument.

From 1895 to 1898 the Austrian physicist on at least three occasions rejected the idea of a very special beginning of the universe, presenting a cosmological scenario where "the whole universe is, and rests forever, in thermal equilibrium" although "here and there" relatively small isolated regions (which he called *Einzelwelten*, i.e.: single worlds, imagining them to have „the extension of our stellar space") can fluctuate temporarily from that state, admitting then the local existence of "visible motion and life"[61].

Assuming the universe was great enough, the fact that a "small part of it as our world should be in its present state" was in complete agreement with the probabilistic interpretations of entropy, since this state was simply very improbable (although not impossible) all the same remaining a preliminary condition for the existence itself of living observers.

---

[57] Tolver Preston, S.: *Temperature Equilibrium in the Universe in Relation to the Kinetic Theory*, Nature 20, 1879, p. 28; Tolver Preston, S.: *On the Possibility of accounting for the Continuance of Recurring Changes in the Universe, consistently with the Tendency to Temperature-Equilibrium*, Philosophical Magazine (5) 8, 1879, p.152/163.

[58] Tolver Preston, S.: *On the possibility of explaining the continuance of life in the Universe consistent with the tendency to temperature-equilibrium*, Nature 19, 1879, p. 460/462 (quotations on p. 462); see also: Tolver Preston, S.: *A Question regarding one of the Physical Premises upon which the Finality of Universal Change is based*, Philosophical Magazine (5) 10, 1880, p. 338/342. An illuminating description of Tolver Preston's scenario was the following: "Thus it appears that the kinetic theory, applied to the universe, would have the peculiar characteristics of allowing almost indefinite local fluctuations of temperature, of states of aggregation, and of composition, of the matter forming the universe within regions very extensive, absolutely speaking (i. e., in comparison with the boundless universe), these regions being amply extensive enough to allow an amount of activity and variability of energy adapted to the conditions of life; while at the same time the principles of the theory, from their very nature, involve perpetually recurring and yet indefinitely variable changes within certain localised limits, the constitution of the vast whole (looked at broadly) remaining uniform throughout" (from p. 462 of Tolver Preston, S.: *On the possibility of… 1879*).

[59] Muir, W.: *Mr. Preston on General Temperature-Equilibrium*, Nature 20, 1879, p. 6.

[60] Fitzgerald as reported in Bryan, G.H: *The kinetic theory of gases*, Nature 51, 1894, p. 152.

[61] Boltzmann, L.: *On certain questions of the theory of gases*, Nature 51, 1895, p. 413-415; Boltzmann, L.: *Zu Hrn. Zermelo's Abhandlung über die mechanische Erklärung irreversibler Vorgänge*, Annalen der Physik 60, p. 392/398; *Vorlesungen über Gastheorie*, Band 2, Barth, Leipzig, 1898. See p. 208 for quotations of the first source and pp. 395 and 257 respectively for the other citations. Note that Boltzmann employed the terms *Fixsternhimmel* and *Sternenraum* as synonyms.

Boltzmann depicted a collection [*Inbegriff*] of single individual worlds reciprocally separated by thermically dead matter along spatial distances $10^{100}$ times larger than the actual distance of Sirius from the Sun (which is about 7 light years). But although such distances are impressively large even if considered in the light of contemporary standards, they still were not infinite (actual infinity, as we will see, is indeed a source of serious problems for probabilistic reasoning).

The re-proposal of the Boltzmann scenario in the context of a spatially infinite universe was anyway a move that Boltzmann's heirs were ready to fulfil, and remains indeed a fundamental step for contemporary multiverse conceptions[62].

Although the problem of the entropy of the universe has passed through drastic changes after the adoption of general relativity, the search for an alternative to very special initial conditions of the whole world remained a major issue in modern cosmology. The Preston/Boltzmann's scenario at the same time did represent a prototype for the elaboration of both inhomogeneous models and (still more) for all those cosmological proposals which imply the co-existence of isolated regions of space/time characterized by peculiar physical properties (possibly conceived as stochastic outcomes of some fundamental process or event as, for instance, phase transitions in the very early universe).

Tolman and Dingle's inhomogeneous models will be discussed below, but let me immediately introduce a few other examples, such as the scenario exposed by Grigory Moiseevich Idlis in a paper of 1958 (entitled *Essential features of the astrophysical observed universe as typical properties of the inhabited cosmic system*[63]) which was considered by Zel'dovich as the first application of the anthropic principle in relativistic cosmology[64].

Openly inspired by Boltzmann, the Russian astronomer presented a cosmological scenario where the observed part of the universe constituted a peculiar region, provided with the properties necessary for the rise, evolution and maintenance of life, contained in an otherwise infinite and eternal universe which is inhomogeneous on a very large scale. He showed, in particular, that any typical inhabited region of the universe would have shared with our observable expanding *metagalaxy* the fact of necessarily being an „isolated" region endowed with „characteristic features" such as an appropriate age, expansion rate, average density, average temperature, and chemical composition.

---

[62] Cf. Treder H.J.: *Boltzmanns Kosmologie und die räumliche und zeitliche Unendlichkeit der Welt*, Wissenschaftliche Zeitschrift der Humboldt-Universität. Jena, Mathematisch-Naturwissenschaftliche Reihe 26, 1977, p. 13/15. Various physicists supported Boltzmann's views before War World II. E.g.: Tatiana Ehrenfest (as testified on p. 1642 of Tolman, R.C.: *On the entropy of the universe as a whole*, Physical Review 37, 1931, p. 1639/1660 and then reported in the preface of Ehrenfest, P.: *The Conceptual Foundations of the Statistical Approach in Mechanics*, Cornell University Press, Ithaca 1959); von Smoluchowski, M.: *Gültigkeitsgrenzen des zweiten Hauptsatzes der Wärmetheorie*, in: *Vorträge über die Kinetische Theorie der Materie und der Elektrizität, gehalten in Göttingen auf Einladung der Kommission der Wolfskehlstiftung, von M. Planck, P. Debye, W. Nernst, M.v. Smoluchowski, A. Sommerfeld und H. Lorentz*, Teubner, Leipzig/Berlin 1914, p. 87/121; Haldane J.B.S.: *The universe and irreversibility*, Nature 122, 1928, p. 808-809; Haldane, J.B.S.: *The Inequality of Man and Other Essays*, Chatto & Windus, London 1932. See also (although he never mentioned Boltzmann): Franklin, W.S.: *On Entropy*, Physical Review 30, 1910, p. 766/775. Among the critics deserve mention: Bronstein, M.P.-Landau, L.: *Über den zweiten Wärmesatz und die Zusammenhangsverhältnisse der Welt in Großen*, Physikalische Zeitschrift der Sowjetunion 4, 1933, p. 114/118, on p. 117; von Weiszäcker, C.F.: *Der zweite Hauptsatz und der Unterschied von Vergangenheit und Zukunft*, Annalen der Physik 5. Folge 36, 1939, p. 275/283. Bronstein and Landau considered Boltzmann's ideas „completely false" because of the implausibility of a fluctuation as large as the visible universe. On the topic see also: Cirkovic, M.: *The thermodynamical arrow of time, Reinterpreting the Boltmann-Schuetz argument*, Foundations of Physics 22, 2003, p. 467/490.

[63] Idlis, G.M.: Основные черты наблюдаемой астрономической Вселенной как характерные свойства обитаемой космической системы; Izvestia Astrofiziceskogo Institute of Kazakhinstan, SSR 7, 1958, p. 39-54. Cf. also my papers quoted on note 10 above. "Boltzmann's cosmogony of quasi-isolated space domains" was revived also in: Treder, H.J.: *Boltzmanns Kosmogonie und die hierarchische Struktur des Kosmos*, Astronomische Nachrichten 297, p. 117-126, 1976. Hans-Jürgen Treder advanced a "mathematical model for a relativistic cosmos with molecular hierarchy" and considered the observed metagalaxy as "one system" within "an infinite manifold of quasi-isolated domains". Moreover, he quoted Selety's papers which will be discussed below. Cf. also: Treder H.J.: *Relativity Theory and Historicity of Physical Systems*, in: *Physics, Logic, and History. Based on the First International Colloquium held at the University of Denver, May 16-20, 1966,* Yourgrau, W.-Breck A.D. (eds.), Plenum Press, New York 1970, p. 253/264 (in part. p. 261).

[64] Zeldovich, Ya. B.: *The birth of a closed universe, and the anthropogenic principle*, Soviet Astronomy Letters 7 (5), p. 322/323. Note that *anthropogenic* is here simply a bad translation for *anthropic*.

Idlis stated that there was no reason for appealing to „anomalous initial conditions" of the whole universe. In fact, maintaining that living beings can observe only regions of the universe that do possess the properties of a typical inhabited system and not any region in the „infinite multiformity" of the universe, he declared that we have no right to globally extrapolate the properties of our observable region.

Probably Boltzmann's view also inspired the "radical departure from the 'steady state' concept" which was advanced by Hoyle and Narlikar in the mid 1960's after the discovery of high energy phenomena due to quasars or radiogalaxies that posed major difficulties to steady state cosmology with continuous creation of matter[65]. Hoyle and Narlikar aimed to prevent an initial singularity and attempted to explain quasars as isolated "pockets of creations" associated to very intense gravitational fields (in other words, as sources for the creation of new matter which act in a discrete way rather than continuously as in the orthodox steady state theory). As a consequence, they supposed that the universe should appear highly inhomogeneous on a very large scale, presenting regions of different average density which are expanding faster than others so as to lower or temporarily halt the creation rate within them.

Hoyle and Narlikar then depicted the existence of bubbles with the size and mass of the observable universe, suggesting that similar bubbles may occur at any place and time, but need not develop synchronously"; finally they argued that we live in a wide but temporary "fluctuation" which can be described as a flat expanding Einstein/de Sitter universe with its own physical phenomenology.

Later, Hoyle insisted on an alterative explanation of the cosmic microwave background founded again on the existence of an infinite and globally stationary universe which is on average much more dense than the observable region (in agreement with the observational facts concerning radiogalaxies). Moreover, he underlined that the fine structure and other fundamental constants could indeed be different in other regions of the universe, remarking consequently the crucial role that "environmental effects" play in "our existence as living beings".

In 1965 Hoyle affirmed that

"the universe would be far richer in its possibilities and content than we normally imagine. In other regions the numbers would be different and the gross properties of matter, the science of chemistry for example, would be entirely changed"[66].

while, eleven years later he imagined that

"the ultimate future of the physical sciences may lie not so much in unravelling the properties of our particular environment as in working through the possibilities permitted by all the kinds of environments that may exist in a universe of far greater complexity than we contemplate in our usual cosmological studies"[67].

In more recent times the Preston/Boltzmann's scenario has then become common occurrence both in eternal inflation[68] and in the landscape which describes the space of all the possible vacua

---

[65] Hoyle, F.-Narlikar, J.V.: *A radical departure from the 'steady state' concept in cosmology*, Proceedings of the Royal Society A290, 1966, p. 162-176.

[66] Hoyle, F.: *Recent developments in cosmology*, Nature 208, 1965, p. 111/114 on p. 114.

[67] Hoyle, F.: *The future of physics and astronomy*, American Scientist 64, 1971 (March-April), p. 197/202 on p. 202.

[68] Eternal inflation, as introduced originally by Andrei Linde, invoked self-reproducing universes in order to avoid a beginning of the universe. Cf.: Linde, A.: *Eternally Existing Self-Reproducing Inflationary Universe*, Physica Scripta T15, p. 169/175, 1987; Linde, A.-Linde, D.-Mezhlumian, A.: *From the big bang theory to the theory of a stationary universe*, Physical Review D49, 1994, p. 1783/1826. Linde's scenario, where the universe as a whole is past-eternal (i.e.: "immortal" in Linde's works), was rejected by Borde and Vilenkin who demonstrated that "a physical reasonable

according to string theory. In the first case, one has a picture of post-inflationary "thermalised islands" (which are multiple regions of space time, each provided of their own low energy constants and physical features) in an "inflating background of false vacuum"; in the latter, one contemplates a scenario of at least $10^{500}$ metastable low-energy vacua produced through several possible inflationary mechanisms (such as those connected with eternal inflation itself) and, of course, representing completely disjointed domains[69].

Both eternal inflation and the landscape scenario imply, therefore, a spatial variation of the cosmological constant (an idea which was introduced in 1931 by the Swiss Gustave Juvet in order to support Shapley's conviction that external spiral nebulae were smaller than the Milky Way, but which remained virtually unknown[70]) and promote an anthropic interpretation of the local value of that "constant" which resembles Boltzmann's treatment.

In all the above mentioned examples (Idlis, Hoyle, eternal inflation and the "landscape" proposal) we can indeed conceive the presence of different universes as separated "islands" in the same "ocean"[71] and evaluate that our presence as observers is possible only in those regions which have the appropriate conditions for the emergence and development of complexity and life, despite the unlikely realization of such conditions.

In a fairly significant way, the picture of "a cosmic archipelago", which is now commonly adopted for describing the "new concept" of the multiverse[72], was previously presented by von Humboldt together with the expression "island universe" (*Weltinsel*)[73] and then adopted by John Pringle Nichol, teacher of lord Kelvin and the main source of Poe's visionary cosmology[74].

## 3.THE CLOSED UNIVERSE AND THE INFINITE WORLDS

The publication of Einstein's *Kosmologische Betrachtungen* in 1917 marks the birth of general relativistic cosmology or, if one prefers, of the modern science of the universe[75]. Although Einstein's main aim was then to find a satisfactory extension of his field equations consistent with his ideas on the nature of inertia (and with the consequent problem of the behaviour of the metric field at spatial infinity), the result of his inquiry brought him to picture a static self-contained universe with closed space sections of constant curvature.

Einstein, in any case, was not the first to apply Riemannian geometry in a cosmological perspective. From Zöllner to Lense, passing from Peirce, Plassmann, Schwarzschild, Barbarin, Harzer,

---

spacetime which is eternal inflating to the future must possess an initial singularity" (Borde, A.-Vilenkin, A.: *Eternal inflation and the initial singularity*, Physical Review Letters 72, 1994, p. 3305/3309). For a review cf.: Guth, A.: *Eternal Inflation*, arXiv: astro-ph/0101507. Note, moreover, that a new attempt to build an eternally inflationary model with no initial time has been advanced in: Aguirre, A.-Gratton, S.: *Steady-State Eternal Inflation*, Physical Review D65, 2002, 083507 or arXiv: astro-ph/0111191; *Inflation without a beginning: a null boundary proposal*, Physical Review D67, 2003, 083515 or arXiv: gr-qc/0301042.

[69] Quotations from Garriga, J.–Vilenkin, A.: *Anthropic prediction for Λ and the Q ...,* 2005. Cf. also: Weinstein, S.: *Anthropic reasoning in multiverse cosmology and string theory*, ArXiv:hep-th/0508006 (August 2005).

[70] Juvet, G.: *Sur quelques solutions des équations cosmologiques de la relativité*, Commentarii Mathematici Helvetici 3, 1931, p. 154/172; *Sur quelques solutions des équations cosmologiques de la relativité - II*, Commentarii Mathematici Helvetici 4, 1932, p. 102/105.

[71] Cf. Hoyle, F.: *Recent developments ...*, 1965, on p. 114.

[72] E.g.: Rees, M.: *Before the ...* , 1997, on p. 4.

[73] von Humboldt, A.: *Kosmos. Entwurf einer physischen Weltbeschreibung*, zweite Auflage, 4 volumes, J. G. Gotta'scher Verlag, Stuttgart 1847.

[74] Nichol, J.P.: *Views of Astronomy*, Greeley & McElrath, New York 1848.

[75] Einstein, A.: *Kosmologische Betrachtungen zur allgemeine Relativitätstheorie*, Sitzungberichte der Königlich Preussischen Akademie der Wissenschaften zu Berlin, phys.-math.Klasse1917, p. 142/152. On Einstein's static model see in particular: Kerszberg, P.: *The Invented Universe. The Einstein-De Sitter Controversy (1916 - 17) and the Rise of Relativistic Cosmology*, Clarendon Press, Oxford 1989; Bettini, S.: *Immagini dell'Universo. Le due genesi della cosmologia*, Ph.D. Thesis, 2001; Röhle, S.: *Mathematische Probleme in der Einstein-de Sitter Kontroverse*, Max Planck Institut für Wissenschaftsgeschichte, preprint 210, Berlin 2000, available on : http://www.mpiwg-berlin.mpg.de/en/forschung/Preprints/P210.PDF.

Frankland and others[76], many authors contributed to make the applications of non-Euclidean geometry to large scale astronomy a rather usual event in the late XIX century[77]. It is not then surprising to find that Alexander Moszkowsky, in his controversial[78] essay published in 1921, discussed along with Einstein the meaning of a Riemannian finite universe before dealing with a topic such as that made popular by Zöllner's and Slade's speculations on an extradimensional world populated by spirits.

In this circumstance, Einstein explained patiently to Moszkowski what other universes should possibly be according to his cosmological model. They were surely not to be conceived as other islands of stars (a possibility that Einstein ignored in his scientific memoirs, figuring a uniform distribution of stars, but that in any case would have not conducted outside the finite universe) but rather as "other universes" without any physical connection with our one ("*andere Universen außer Zusammenhang mit diesem*") and "outside from whatever investigable connection" (*außer einem jemals erforschbaren Zusammenhang*)[79].

Einstein's description of such external *Ultra Welten* was not dissimilar from many popular expositions of non Euclidean Geometry of his day. In fact, Einstein explained that the conception of "another universe which proceeds parallel" (*ein anderes parallel verlaufendes Universum*) implied a "cosmic plan" whose "phenomena and relations" were destined to remain forever inaccessible to experiment and observations. In other words, he suggested that we were in the same position of the inhabitants of a two-dimensional Flatland which would have ignored forever the possible presence of a two dimensional world completely separated from theirs[80].

A similar picture had clearly a pedagogical function. In any case, it is not so conceptually different from all those proposals that will illustrate the co-existence of "universes" in some sort of superspace (as the Hilbert space invoked by the many worlds formulation of quantum mechanics or

---

[76] Zöllner, J.C.F.: *Über die Natur der Cometen: Beiträge zur Geschichte und Theorie der Erkenntniss*, Wilhelm Engelmann, Leipzig 1872; Peirce, C.S.: *The Architecture of Theories*, The Monist 1, 1891, p. 161/176; Plassmann, J.: *Himmelskunde: Versuch einer methodischen Einführung in die Hauptlehren der Astronomie*, Herdersche Verlagsbuchhandlung, Freiburg i. Br. 1898; Schwarzschild, K.: *Ueber das zulässige Krümmungsmass des Raumes*, Vierteljahrschrift der astronomischen Gesellschaft 35, 1900, p. 337/347; Barbarin, P.: *Sur le paramètre de l'univers*, Mémoires de la Société des sciences physiques et naturelles de Bordeaux 1900/1901, p. 71/74; Harzer, P.: *Die Sterne und der Raum*, Jahresbericht der Deutschen Mathematiker-Vereinigung 1908, p. 237/267; Frankland, W. B.: *Notes on the Parallel Axiom*, The Mathematical Gazette 7, 1913, p. 136/139; Lense, J.: *Das Newtonsche Gesetz in nichteuklidischen Räumen*, Astronomische Nachrinten 205, 1917, p. 241/248; Lense, J.: *Das Newtonsche Gesetz in nichteuklidischen Räumen*, Sitzungsberichte der Akademie der Wissenschafte. Mathematisch-naturwissenschaftliche Klasse 126, 1917, p. 1037/1063. Simon Newcomb discussed the possibility of a positively curved space in: Newcomb, S.: *Popular Astronomy*, ... 1878; Newcomb, S.: *The philosophy of hyperspace*, Bulletin of the American Mathematical Society (2) 4, 1898, p. 187/195. Riemann, on the other hand, remarked that speculations about the " incommensurably great" [*Unmessbargrosse*] were idle [*müssig*] (Riemann, B.: *Über die Hypothesen, welche der Geometrie zu Grunde liegen*, Abhandlungen der Königlichen Gesellschaften der Wissenschaften zu Göttingen 13, 1867, now available on. http://www.maths.tcd.ie/pub/HistMath/People/Riemann/Geom/ ) while Clifford stated that the geometer knew "nothing about the nature of actually existing space at an infinite distance" (Clifford, W.K.: *The Postulate of the Science of Space*, 1873, now available in: http://www.math.uwaterloo.ca/navigation/ideas/articles/clifford/index.shtml ). Finally, Maxwell rejected any non-Euclidean conceptions attributing them to "space crumplers" (postcard to Tait of 1874, reproduced in Harman, P.H.: *Energy, Force, and Matter. The Conceptual Development of Nineteenth-Century Physics*, Cambridge University Press, *Cambridge* 1982).

[77] Cf.: Sommerville, D.M.Y.: *Bibliography of non-euclidean Geometry*, University of St. Andrews, Scotland and Harrison & Sons, London 1911; North, J.D.: *The Measure of the Universe. A History of Modern Cosmology*, Oxford University Press, Oxford 1965. Debates on non-Euclidean geometries and curved space often made references also to extra-dimensions of space. On this last topic see: Bettini, S.: Immagini dell'Universo ..., 2001, part II, section 4.8 and references quoted therein.

[78] Cf. Clark, R.W.: *Einstein, the Life and Times*, World Publishing, New York 1973.

[79] Moszkowski, A. 1921: *Einstein. Einblicke in seine Gedankenwelt. Gemeinverständliche Betrachtungen über die Relativitätstheorie und ein neues Weltsystem. Entwickelt aus Gesprächen mit Einstein*, F. Fontane & Co., Berlin 1921; quotations on p. 133.

[80] Compare Einstein's example with the following remark by Newcomb: "Right around us, but in a direction which we cannot conceive any more than the inhabitants of 'flat-land' can conceive up and down, there may exist not merely another universe, but any number of universes" (Newcomb, S.: *Side-Lights…, 1906).

jointly with Carter's strong anthropic principle), by Tegmark's level 3 parallel universes[81] or by cosmological models as that proposed by Jaroslav Pachner in 1960. This last author suggested that

"From the finite radium of the universe, we may not conclude that it is the only existing one. We should suppose the existence of many closed universes so embedded into the "cosmical space of higher number of dimensions than four" that their hypersurfaces do not intersect each other. Since there exists no physical interaction under them, they are incapable of being observed, but this does not signify that they do not exist"[82].

Despite these future developments, Moszkowsky's collection of Einstein's thoughts represents probably nothing more than a simple curiosity although it maintains a certain fascination if one thinks of the controversies which were raised by the idea of a curved finite universe and associated with the relativistic view of the world.
Only de Sitter, Weyl, Felix Klein, Eddington, Lanczos and a few other scholars with competence in tensor calculus and general relativity were able to participate directly in that abstruse debate which followed the publication of Einstein's *Kosmologische Betrachtungen*. Many scientists, however, felt the necessity to reject the idea itself of a finite universe in the name of the infinity of the world.
Amongst them were the partisans of an infinite hierarchical universe composed of "a boundless vista of worlds within worlds"[83].
Hierarchic cosmology and the scenario of a succession of worlds within worlds which proceed indefinitely both in the direction of the infinitely large and of the infinitely small has a long history, which can be traced back to Pascal's thought on the disproportion of man (which contemplated the image of "an infinity of universes" each with "its own firmament, planets, earth, in the same proportion as the visible world"[84]) and to Leibniz metaphysics. The hierarchical conception of the system of the world knew then a notable diffusion with Lambert's *Cosmologische Briefe*, an essay which was translated in English, French and Russian but which does not appreciate in any way the idea of an infinite world[85].
On the other hand, it was in order to face the problem of infinity that the historian of atomism Kurd Lasswitz presented, in 1877, his version of the hierachic scenario[86]. He suggested that the incomprehensibility of the concept of infinity was due to an incapacity of our cognitive faculty (*unserer Erkenntnissfähigkeit*) and discussed in depth the so called "relativity of magnitude".
The idea according to which "lengths, times and masses are reciprocally proportional in systems of different order, whereas relative velocities are the same"[87] was truly fundamental in the formulation of Edmund Edward Fournier d'Albe's version of hierarchic cosmology which was first presented in chapter XVI of *The Electron Theory* (1906) and, one year later, in a celebrated essay entitled *Two New Worlds: The Infra World. The Supra World*[88].

---

[81] Tegmark, M.: *Parallel Universes*, in: *Science and Ultimate Reality: From Quantum to Cosmos, honoring John Wheeler's 90th birthday*, Barrow, J.D.- Davies, P.C.W.-Harper, C.L. (eds.) Cambridge University Press, Cambridge 2003, or arXiv:astro-ph/0302131.
[82] Pachner, J.: *Dynamics of the universe*, Acta Physica Polonica 19, 1960, p. 663/673, on p. 673.
[83] Fournier d'Albe, E.E.: *The Electron Theory. A popular introduction to the new theory of electricity and magnetism*, Longmans Green & Co., London 1906, on p. 290.
[84] B. Pascal, *Pensées*, published in 1669, translation from: http://oregonstate.edu/instruct/phl302/texts/pascal/pensees-contents.html (Thought n. 72).
[85] Lambert, J.H.: *Cosmologische Briefe über die Einrichtung des Weltbaues*, E. Kletts Wittib, Augsburg 1761; English translation (due to S.L. Jaki): *Cosmological Letters on the Arrangement of the World-Edifice*, Scottish Academic Press, Edinburgh 1976.
[86] Lasswitz, K.: *Ein Betrag zum kosmologischen Problem und zur Festellung des Unendlichkeitsbegriffes*, Vierteljahrsschrift für wissenschaftliche Philosophie 3, Leipzig 1877, p. 329/360.
[87] Selety, F.: *Possibilité d'un potentiel infini, et d'une vitesse moyenne de toutes l'ètoiles ègale à celle de la lumière*, Compte Rendu Hebdomadaire des Scéances de l'Académie des Sciences 177, 1923, p. 250/252, on p. 250.
[88] Fournier d'Albe, E.E.: 1907: *Two New Worlds. The Infra World. The Supra World*, Longmans Green & Co., London 1907.

Inspired by the "chemico-astronomical" analogy between the current models of the atom and the Solar system, Fournier imagined a hierarchical universe where each level was $10^{22}$ times (i.e.: approximately the ratio between the diameter of the electron and that of the Earth) larger than the preceding one and consequently put forward a peculiar solution of Olbers' paradox. Such a proposal was greatly appreciated by the Swedish astronomer Carl Vilhelm Ludwig Charlier who defined Fournier's book of 1907 "ingenious" and, a year later, elaborated a more cogent mathematical model for an infinite hierachical world not afflicted by Olbers' and Seeliger's paradoxes[89].

Charlier further elaborated that model after the publication of Einstein's cosmological considerations animated by the defense of "an infinitely extended world" and by the need to establish a quantitative "doctrine of infinite rarity of matter in space"[90]. In 1922, he advanced also a natural explanation for the observed recessional velocities of spiral nebulae.

The scenario depicted by Charlier, where stars form Milky Ways, Milky Ways form a system of a higher order and so on indefinitely, was sometimes described as a sequence of "super-universes"[91] and gave rise to an extreme form of steady state cosmology. In Charlier's model, in fact, the universe appears the same not only in any place and at any time but also on any scale (and life, moreover, can be present at any level of the hierarchy).

Einstein, at any rate, did not debate directly with Charlier but with the Austrian philosopher Franz Josef Jeiteles who, after having changed his name to Franz Selety in 1918, presented in 1922 a further elaboration of the hierarchical cosmology in the form of a "world with molecular hierarchy" [*molekularhierarchische Welt*][92].

Contrary to his predecessors, Selety paid genuine attention to the topics discussed in the circle of general relativistic experts and aimed not only at the construction of a non contradictory infinite Newtonian universe but also at the setting-up of an infinite hierarchical model compatible with general relativity and conceivable as an alternative to Einstein's finite universe. Incidentally, he adopted a pretty modern approach for his times, suggesting that cosmology had to pay attention not only to the structure of the actual world but also to the mathematically possible universes.

Selety was indeed the first to conceive (two years before Friedmann's second memoir) a consistent general relativistic open model with Euclidean space sections.

In September 1922 Einstein dedicated one of his last interventions on cosmology of the 1920's to reject Selety's thesis[93]. He accepted the hierarchical scenario as a way out from the paradoxes of Newtonian's cosmology but then attacked both Charlier and Selety in a very unfortunate way, i.e. denying the empirical evidence of external galaxies. Moreover, Einstein affirmed that a world with molecular hierarchy was compatible with general relativity but unsatisfying because unable to substantiate a material determination of inertia. At last the inventor of general relativity repeated his arguments against an infinite universe already discussed in his Princeton's lectures some months before[94]. In short: a closed surface is preferable to an open one on the ground of simplicity[95];

---

[89] Charlier, C.V.L.: *Wie eine unendliche Welt aufgebaut sein kann*, Arkiv för Matematik, Astronomi och Fysik 4, 1908, p. 1/15, quotation on p. 1.

[90] Charlier, C.V.L.: *How an infinite world may be built up*, Arkiv för Matematik, Astronomi och Fysik 16, n. 22, 1922, p. 1/34. Quotations are from the review of Charlier's cosmology given in: Arrhenius, S.: *Die Unendlichkeit der Welt*, Scientia 5, 1909, p. 217/229. The second of Charlier's paper gained a wide popularity among his contemporaries; cf.: Bernheimer, W.E.: *C. V. L. Charliers Untersuchungen über den Aufbau einer unendlichen Welt*, Naturwissenschaften 20, 1922, p. 481/484; Smart, W.M.: *An Infinite Universe*, Observatory 45, 1922, p. 216/222; Silberstein, L.: *The Theory of Relativity*, second edition, Macmillan & Co., London 1924; Russell, B.: *The ABC of Relativity*, first edition, Kegan Paul, Trench, Trubner & Co., London 1925.

[91] Thomas, O.: *Himmel und Welt*, Arbeitsgemeinschaft für Kultur und Aufbau, München 1928. Quotation on p. 228 of the English translation due to B. Miall: *The Heavens and the Universe*, George Allen & Unwin, London 1930.

[92] Selety, F.: *Beiträge zum kosmologischen Problem,* Annalen der Physik 68, 1922, p. 281/334.

[93] Eintstein, A.: *Bemerkung zu der Franz Seletyschen Arbeit "Beitrage zum kosmologischen System"*, Annalen der Physik 69, 1922, p. 436/438.

[94] Einstein, A.: *Vier Vorlesungen über Relativitätstheorie*, Vieweg, Braunschweig 1922; English translation (due to E.P. Adams): *The Meaning of Relativity: Four Lectures delivered at Princeton University, May 1921*, Princeton University Press, Princeton 1921.

Mach's principle implies a spatially closed universe; the hypothesis of a finite average density of matter is preferable to the request of an average density tending to 0.

Selety disagreed with all these three arguments[96]. He rejected vigorously the conclusion according to which a Riemannian world was simpler than a Minkowskian one and defined Einstein's interpretation of Mach's ideas just a mere "opinion". In favour of the infinity of the universe he added then the concept of relativity of length; the possibility of a "cosmological satisfaction" of the principle of special relativity in a Minkowskian context and the observation that an irregular distribution of matter was closer to the actual facts than the uniform distribution assumed in Einstein's cosmology.

In any case, Selety was one of the few defenders of an infinite universe who discussed this topic directly with Einstein. The large majority of those who invoked the infinity of the universe (from Arrhenius to Nernst, passing fromWiechert, MacMillan and Charlier) simply cultivated a sort of spontaneous hostility towards general relativity and its model building claims[97].

Amongst them was Emile Borel who, apart from disclosing a certain sympathy for Charlier's cosmology[98], proposed on a couple of occasions a very strong argument against both Einstenian cosmology and any other attempt to build a model of the whole universe[99]. The French mathematician argued in fact that, in a truly cosmological perspective, we are in the same position as tiny inhabitants of a drop of water: just as they would be unable to appreciate the complexity of the world outside the drop, we would analogously be completely on the wrong track in trying to extrapolate conclusions from the "visible universe" to the whole totality of things.

Borel was surely inspired by Boltzmann's approach to cosmology in stating that "local knowledge can't give knowledge of the universe"[100] and his conclusions represented a warning for any theoretical attempt aimed to extrapolate beyond what is observed from our spatio/temporal collocation (or, in other words, what is outside of our past light cone). Cosmology, as also noted recently in several occasions, cannot avoid recourse to unverifiable assumptions, while the adoption of principles of different kinds can indeed lead to very different conclusions and to incompatible scenarios[101].

In any case, the possibility of a hierarchical universe (i.e. a cosmology which does not accept the cosmological principle and presupposes an infinite space with null average density) was revived from time to time in the history of modern cosmology[102]. As an alternative to the homogeneous and isotropic expanding models, it represented a topic often discussed by Soviet astronomers[103], strongly supported by Gerard de Vaucouleurs in the 1970's and brought back in vogue again in the form of the so called "fractal cosmology". Recent proposals cannot, in any case, maintain the concept of relativity of magnitude or Fournier's scenario of worlds within worlds. The idea of universes within atoms is clearly untenable in the light of contemporary physics[104], although one could be tempted to make a parallel with those recent speculations suggestimg that black holes are the cradles of new universes provided with their own physical phenomenology[105].

---

[95] Einstein justified this assertion saying that spatial infinity involves a fourth rank tensor (i.e. the Riemann's tensor) which vanishes at infinity, whereas the gravitational filed of a closed space does not require more than a second rank tensor (cf. Einstein, A.: *Vier Vorlesungen*..., p. 64).

[96] Selety, F.: *Unendlichkeit des Raumes und allgemeine Relativitätstheorie*, Annalen der Physik 73, p. 291/325. Cf. also: Selety, F.: *Erwiderung auf die Bemerkungen Einsteins über meine Arbeit "Beiträge zum kosmologischen Problem"*, Annalen der Physik 72, p. 58/66.

[97] Cf. Kragh, H.: *Cosmology between the wars: the Nernst-MacMillan alternative*, Journal for the History of Astronomy 26, 1995, p. 93/115 and the special issue of the Canadian journal Apeiron. Studies in infinite nature, dedicated to Nernst's cosmology (vol. 2, no. 3, July 1995, avalaible on http://redshift.vif.com/journal_archives.htm ). To appreciate Charlier's view cf. also Charlier, C.V.L.: *On the Structure of the Universe*, Publications of the Astronomical Society of the Pacifics 37, 1925, p. 177/191.

[98] Cf.: Borel, E.: *Dèfinition arithmètique d'une distribution de masses s'ètendant à l'infini et quasi periodique avec une densitè moyenne nulle*, Compte Rendu Hebdomadaire des Scéances de l'Académie des Sciences 174, 1922, p. 977/979; *L'espace et le temps*, F. Alcan, Paris 1922, English translation (due to A.S. Rappoport and J. Dougall): *Space and Time*, Dover, New York 1960.

[99] Preface dated November *1920* of the French edition of: A. Einstein, *La théorie de la relativité restreinte et généralisée*, 1921; reprinted in: *Œuvres de Emile Borel*; tome III, Editions du Centre National de la Recherche Scientifique, Paris 1972, p. 1839/1853 (on p. 1844); Borel, E.: *L'espace et le temps* …, 1922, p. 227 of the English translation. Cf. also: Richard-Foy, E.: *Einstein et sa conception d'un espace fini*, Revue Philosophique de la France et de l'Etranger 97, 1924 p. 67/103, in part. p.85.

[100] Cf. Borel, E.: *L'espace et le temps* ..., 1922, p. 114 of the English translation.

[101] Cf. Ellis, G.F.R.: *Cosmology and Verifiability*, Quarterly Journal of the Royal Astronomical Society 16, 1975, p. 245/264; *Limits to Verification in Cosmology*, Annals of the New York Academy of Sciences 336, 1980, p. 130/160; *Cosmology: Observational Verification, Certainty and Uncertainty*, South African Journal of Science 76, 1980, p. 504/511.

The topic of actual infinity is, instead, a recurring source for concern. It is, in fact, still common to find in literature a restatement of Kepler's objection to the actual infinite (and of Bruno's infinite worlds) according to which the existence of infinitely many planets would imply the presence of "as many Galileos observing new stars in new worlds as there are worlds". The existence of *doppelgänger* was indeed frequently discussed in the XIX and early XX centuries. It was, for instance, cited by the socialist theorist and anti-positivist thinker Louis-Auguste Blanqui in an essay of 1872, *L'éternité par les astres,* which was later appreciated by both Benjamin and Borges.

Blanqui reflected on the consequences of having an infinite number of stars and a finite number of "simple bodies" (i.e. our chemical elements[106]) and concluded that:

"The number of our doubles is infinite in time and space. In all conscience, we can hardly ask for more. These doubles are of flesh and blood, or in pants and coats, in crinoline and chignon. These aren't phantoms: they are the now eternalized. There is nevertheless a great defect: there is, alas, no progress! No, these are vulgar re-editions, repetitions. As it is with editions of past worlds, so it is with those of future worlds. Only the chapter of bifurcations remains open to hope. Never forget that *all we could have been here, we are somewhere else"[107]*.

Similar conclusions were criticized by von Nägeli[108], discussed by Engels[109] and revived then in the debates on Nietzsche's eternal return and Poincaré's theorem[110]. Finally, they represented a subject for cosmological speculations[111]. The point is that Blanqui's problematic insight seems to maintain its validity also when one considers the open expanding universes of relativistic cosmology or the spatially infinite domains of some inflationary scenarios. This is at least what was claimed to have been demonstrated by Ellis and Brundrit for open isotropic and homogeneous expanding universes[112] and, more recently, by Garriga/Vilenkin and Knobe/Olum/Vilenkin in the case of inflationary models[113]. Barrow and Tipler have on their part remarked that an infinite repetition as that imagined by Blanqui requires not only an infinite space but also "an exhaustively random infinity in order to include all the possibilities"[114]. Despite this last warning[115], it remains a rather common occurrence to read that an infinite space is by itself sufficient for having replicas of ourselves somewhere in the universe[116]; while the duplication becomes that of whole universes (or better still: of an infinite number of regions "whose history is identical to ours") once one considers, as Garriga and Vilenkin do, that "all histories consistent with exact conservation laws will have non-vanishing probabilities and will occur in an infinite number" of domains within the thermalised regions of certain inflationary models[117].

---

[102] E.g.: Wilson, A.C.: *A Hierarchical Cosmological Model*, Astronomical Journal 72, 1967, p. 326/327; De Vacouleurs, G.: *The Case for a Hierarchical Cosmology*, Science 167, 1970, p. 1203/1213; Wertz, J. R.: *A Newtonian Big-Bang Hierarchical Cosmological Model*, Astrophysical Journal 164, 1971, p. 227/236; Bonnor, W.B.: *A Non-uniform Relativistic Cosmological Model*, Monthly Notices of the Royal Astronomical Society 159, 1972, p. 261/268; Mandelbrot, B.B.: *Les objects fractal*, Flammarion, Paris 1975; Wesson, P.S.: *General-relativistic hierarchical cosmology*: *an exact model*, Astrophysics and Space Science 54, 1978, p. 489/495*;* Oldershaw, R.L.: *Hierarchical Cosmology*, Astrophysics and Space Science 189, 1992, p. 163/168.

[103] Cf. Kazutinski, V. 1971: *Etat actuel de la théorie cosmogonique*, in: *Problémes de cosmogonie contemporaine*, Ambartsoumian, V.A. (ed.) Mir, Moscow 1971, p. 333/370 and references quoted therein. Idlis himself dedicated a paper to hierarchical cosmology: Idlis, G.M.: Теория относительности и структурная бесконечность Вселенной [The theory of relativity and the structural infinity of the universe], *Astronomicheskii Zhurnal* 33, 1956, p. 622/626.

[104] In any case Erasmo Recami and collaborators still speak of hadrons as micro-universes within a theoretical context (the bi-scale theory of gravitational and strong interactions) which contemplates the hierarchical scenario and explicitly recalls the contributions of Charlier and Selety. E.g.: Recami, E.-Ammiraju, P.-Hernández, H.E.-Kretly, L.C.-Rodrigues jr. W.A.: *Elementary particles as micro-universes. A geometrical approach to strong gravity*, Apeiron 4, n.1, 1997, p. 7/16.

[105] Smolin, L.: *Did the universe evolve?*, Classical and Quantum Gravity 9, 1992, p. 173/191; *The Life of the Cosmos*, Oxford University Press, Oxford 1997. Cf.: Vaas, R.: *Is there a Darwinian Evolution of the Cosmos? - Some Comments on Lee Smolin's Theory of the Origin of Universes by Means of Natural Selection*, arXiv:gr-qc/0205119; Smith, Q.: *The black hole origin of the Universe: frontiers of speculative, current physical cosmology*, paper presented at the Internal Conference on Physical Cosmology, Santa Barbara, 2000, in: http://www.qsmithwmu.com/ , 2004.

[106] Blanqui suggested the existence of no more than 100 simple bodies.

[107] Blanqui, L.A.: *L'éternité par les astres*, Libraire Germer Baillière, Paris 1872, English translation as in http://www.marxists.org/reference/archive/blanqui/1872/astres.htm .

[108] von Nägeli, K.W.: *Die Schranken der naturwissenschaften Erkenntnis*, 1877, reprinted. in: von Nägeli, K.W.: *Mechanisch-physiologische Theorie der Abstammungslehere*, R. Oldenburg, München 1884, p. 555/602.

[109] Engels, F.: *Dialektik der Natur*, Dietz Verlag, Berlin 1952 (the book collects papers and notes dated 1877).

---

[110] Cf. Brush, S.G.: *The Kind of Motion We Call Heat. A History of the Kinetic Theory of Gases in the 19th Century*, 2 volumes, North Holland, Amsterdam 1976; D'Iorio, P.: *La linea e il circolo. Cosmologia e filosofia dell'eterno ritorno in Nietzsche*, Pantograf, Genova 1995. Nietzsche probably read Blanqui's book (cf. Andina, T.: *Eterno Ritorno: Nietzsche, Blanqui e la cosmologia del Big Bang*, Rivista di Estetica 41, 2001, p. 3/35).

[111] E.g.: Selety, F.: *Über die Wiederholung des Gleichen im kosmischen Geschehen, infolge des psychologischen Gesetzes der Schwelle*, Zeitschrift für Philosophie und philosophische Kritik 155, 1914, p. 185/205. Cf. Selety's letter to Einstein of July 23, 1917, in: *The Collected Papers of Albert Einstein. Volume 8: The Berlin Years: Correspondence, 1914 - 1918*, Schulmann, R.-Kox A. J.-Janssen, M.-Illy, J. (eds.), Princeton University Press, Princeton 1998, p. 486/494, on p. 494.

[112] Ellis, G. F. R.–Brundrit, G. B. 1986: *Life in the Infinite Universe*, Quarterly Journal of the Royal astronomical Society 20, p. 37/41. Brundrit and Ellis noted that with: -an infinite number of galaxies -a finite probability of intelligent life evolving on other planets and -a finite number of different configurations of the DNA molecule, "there will be an *infinite* number of genetically identical beings in the Universe at any time". At any rate, they remarked in a footnote that because an event can "occur in infinitely many planetary systems", initial conditions must be "such that the probability of *any* event occurring is not vanishing". The impression is that Ellis invoked the possibility of infinite duplication in order to reject spatial infinity. In other papers Ellis resumes the old argument of Christ's multiple incarnations (Ellis, G.F.R.: *The Theology of the Anthropic Priniple*, in: *Quantum Cosmology and the Laws of Nature: Scientific Perspectives on Divine Action*, Russell, R.J.-Murphy, N.-Isham, C.J. (eds.), Vatican Observatory 1993, p. 363/400) and suggests that we are living in a small closed universe with non trivial topology. See: Ellis, G. F. R./Schreiber, G.: *Observational and dynamic properties of small universes*, Physics Letters A115, 1986, p. 97/107; Ellis, G.F.R.: *Observational Properties of Small Universes*, in: *Theory and Observational Limits in Cosmology, Proceedings of the Vatican Observatory Conference Held in Castel Gandolfo July 1 - 9, 1985*, Specola Vaticana, Stoeger, W.R. (ed.) 1987, p. 475/486 . Ellis' small universe was refuted on observational grounds after Oliveira Costa, A.-Smoot, G.F.: *Constraints on the Topology of the Universe from the 2 Tear COBE Data*, Astrophysical Journal 448, 1995, p. 477/481. In any case, the possibility of a small universe with dodecahedral spherical topology has been reconsidered after recent analysis of cosmic microwave background. Cf.: Luminet, J.P.-Weeks, J.R-Riazuelo, A.-Lehoucq, R.-Uzan, J.P.: *Dodecahedral space topology as an explanation for weak wide-angle temperature correlations in the cosmic microwave background*, Nature 425, 2003, p. 593/595. Cf.: Luminet, J.P.: *The Shape of Space after WMAP data*, arXiv: astro-ph/0501189.

[113] Garriga, J.–Vilenkin, A.: *Many worlds in one*, Physical Review D64, 2001, 043511 or arXiv: gr-qc/0102010; Knobe, J.–Olum, K.D.- Vilenkin, A.: *Philosophical Implications of Inflationary Cosmology*, ArXiv:physics/0302071, 2003.

[114] Barrow, J.D.-Tipler, F.J: *The Anthropic Cosmological ...*, 1986, on p. 24.

[115] Cf. Mosterin, J.: *Anthropic Explanations in Cosmology*, To appear in: *Proceedings of the 12th International Congress of Logic, Methodology and Philosophy of Science. Amsterdam,* Hajek-Valdés- Westerstahl (eds.), North-Holland Publishing, Amsterdam 2004 or on: http://philsci-archive.pitt.edu/archive/00001609/ . See also: Coule, D.H.: *Is there paradox with infinite space?*, arXiv:gr-qc/0311022 (February 2004).

[116] E.g.: Tegmark, M.: *Parallel Universes*, ..., 2003; Rees, M., *Our Cosmic Habitat*, Princeton University Press, Princeton 2001. cf. also Tegmark, M.: *Max' multiverse FAQ* in http://space.mit.edu/home/tegmark/multiverse.html .

[117] Garriga, J-Vilenkin, A.: *Many Worlds ...*, 2001.

All this (and the conception itself according to which "anything that can happen will happen; in fact, it will happen an infinite number of times" which was championed by Sciama[118]) is philosophically disturbing for all those that still tend to resuscitate all the classical arguments elaborated to deny actual infinity from Aristotle to Hilbert. The crux of the matter remains the physical actual realization of an infinite universe with an infinite amount of mass/energy[119].

Expanding universes with constant negative curvature (which means infinite space sections if one assumes trivial topology[120]) were discussed for the first time by Friedmann[121] in 1924 who attributed them an "exceptional interest" (probably of the same grade, if not greater, of that stimulated by the non-stationary solutions of the field equations themselves)[122]. Paradoxically enough, the open expanding model with Euclidean space sections became then a fundamental tool of cosmological enquiry thanks to a joint paper by Einstein and de Sitter[123].

Such a model, the so called Einstein/de Sitter universe, was conceived when the two founders of general relativistic cosmology met each other in Pasadena on January 1932 and was then presented as a description of the relation between the expansion and the average density which could result theoretically useful "for the sake of simplicity".

Effectively, the monotonically expanding flat model with null curvature, average density always[124] equal to critical density ($\Omega = 1$) and no cosmological constant, surely represented "the simplest of all known universes"[125]. This despite the fact that one of its authors had declared just some years before to be almost sure that an Euclidean universe was too complex to be acceptable and that, furthermore, neither Einstein nor de Sitter ever lent much relevance to the 1932 paper[126].

Notwithstanding the 1932 paper, Einstein remained responsible for a widespread predilection of the closed universe; a predilection which was shared by many experts in general relativity in the years to come (such as Wheeler[127] or Infeld, with the latter ready to consider Einstein/de Sitter's model rough and unattractive[128]) and that urged others to discard any infinite universe as beyond the conception of the mathematicians themselves[129] if not as "a scandal to human thought"[130].

---

[118] Sciama, D.: *The Anthropic Principle and the non-uniqueness of the Universe*, in: *The Anthropic Principle: Proceedings of the second Venice Conference on Cosmology and Philosophy*, Bertola, F.-Curi. U. (eds), Cambridge University Press, Cambridge 1993, p. 107/110; *Is the Universe unique?*, SISSA preprint ILAS/LL - 17/1995.

[119] Whitrow, G.J.: *The Natural Philosophy of Time*, Nelson, London 1961; Huby, P.M.: *Kant or Cantor? That the universe, if real, must be finite in both space and time,* Philosophy 46, 1971, p. 121/132; Stoeger, W.R.-Ellis, G.F.R.–Kirchner, U.: *Multiverses and Cosmology: Philosophical...*, 2004; Stoeger, W.R.: *What is the 'Universe' which Cosmology Studies?*, in: *Fifty Years in Science and Religion. Ian G. Barbour and his Legacy,* Russell, J.R. (ed.), Ashgate, Oxford 2004, p. 127/143.

[120] Curiously, Friedmann made a serious mistake suggesting in his semi-popular essay of 1923 that universes with positive constant curvature were not necessarily finite. On the contrary, spherical and elliptical spaces are necessarily topological compacts. See Friedmann, A.: *Mir kak prostranstvo i vremya*, Akademiya, Petrograd 1923. French translation (due to J. P. Luminet and A. Grib): *L'univers comme espace et temps*, in: *Alexandre Friedmann Georges Lemaitre. "Essais de cosmologie"*, Luminet, J.P./Grib, A. (eds.) Editions du Seuil, Paris 1997, 1997, p. 99/213.

[121] Friedmann,A.: *Über die Möglichkeit einer Welt mit konstanter negativer Krümmung des Raumes*, Zeitschrift für Physik 21, 1924, p. 326/332. The first to reconsider expanding models with negative curvature in the 1930's was Heckmann (Heckmann, O. H. L.: *Über die Metrik des sich ausdehnenden Universums*, Nachrichten von der Gesellschaft der Wissenschaften zu Göttingen 1931, p. 126/130).

[122] Cf. Tropp, E. A.-Frenkel, V.Ya.-Chernin, A. D.: *Alexander A. Friedmann: the Man who Made the Universe Expand*, Cambridge University Press, Cambridge 1993, p. 170/171.

[123] Einstein, A.-de Sitter, W.: *On the Relation between the Expansion and the Mean Density of the Universe*, Proceedings of the National Academy of Sciences of the U.S.A. 18, 1932, p. 213/214.

[124] Except of course for time zero, where it is undefined.

[125] Harrison, E.R.: *Cosmology. The Science of the Universe*, Cambridge University Press, Cambridge 1981, p. 286.

[126] As apparently testified in: Eddington, A.S.: *Forty years of astronomy*, in: *Background to Modern Science: Ten lectures at Cambridge arranged by the History of Science Committee*, 1936, Needham, J.-Pagel, W. (eds.): Cambridge University Press, Cambridge 1938, p. 115/142, on p. 218.

[127] E.g.: Wheeler , J.A.: *Einstein's Vision*, Springer, Berlin 1968.

[128] Cf. Infeld, L. 1949: *General Relativity and the Structure of Our Universe*, in: *Albert Einstein: Philosopher-Scientist*, Schilpp, P.A. (ed.), Library of Living Philosophers, Evanston (Illinois) 1949, p. 333/355.

[129] Cf. Eddington, A.S.: *Can gravitation be explained?*, Scientia 33, 1923, p. 313/324, on p. 316.

It was indeed only after 1974 (with the publication of an important paper by Beatrice Tinsley, Richard Gott, Jim Gunn and David Schramm) that an open universe became favoured on observational grounds[131]. It was then a consequence of the favour bestowed on inflationary models if the limiting case with Euclidean space sections addressed by Einstein and de Sitter assumed an almost paradigmatic role in recent researches.

## 4. BUBBLES, CYCLES AND POSSIBLE UNIVERSES

After 1930, with the general acceptance of relativistic expanding models, following Eddington's re-discovery of Lemaitre's work, scholars became acquainted with the many possible homogeneous and isotropic expanding models. From 1932/1933 many reviews[132] were dedicated to the different possible dynamical behaviour associated with the line element of non static space/times of constant curvature which satisfy those assumptions which will soon be called "the cosmological principle"[133] (or, preferably, those universes which obey the simplified version of Einstein's equations known as Friedmann's equations). The best known review was that of Howard Percy Robertson dedicated in 1933 to what he called the "Friedmann universes"[134] and what we here, in homage to the main contributors of an intricate history, prefer to call the Friedmann/Lemaitre/Robertson/Walker (in short: FLRW) universes.

As Jacques Merleau-Ponty once brilliantly wrote, all this (a path that culminated in the fundamental contributions of Robertson and Arthur Walker of the mid 1930's[135]) constituted a step from "uncertain intuitions" to a "well defined axiomatic"[136].

At any rate, although the study of the dependence of such parameters as the average density, pressure or the cosmological constant on the scale factor (in the different cases of positive, negative or null curvature) soon became very general, not all the researchers accepted to assign a physical significance to the whole collection of possible FLRW universes.

---

[130] Barnes, E.W. in: *Contributions to a British Discussion on the Evolution of the Universe*, Nature 128, 1931, p. 719/722 on p. 720; *Scientific Theory and Religion*, Cambridge University Press, Cambridge 1933, p. 392.

[131] Gott III, J. R.-Gunn, J.E.-Schramm, D.N.-Tinsley, B.M.: *An Unbound Universe?*, Astrophysical Journal 194, 1974, p. 543/553.

[132] E.g.: de Sitter, W.: *On the expanding universe*, Koninklijke Akademie van Wetenschappen te Amsterdam. Section of Sciences, Proceedings 35, 1932, p. 596/607; *The astronomical aspects of the theory of relativity*, University of California Publication in Mathematics 2, n. 8, Berkeley 1933, p. 143-196; Juvet, G.: *Sur quelques solutions ...*, 1931; Heckmann, O.H.L.: *Die Ausdehnung der Welt in ihrer Abhängigkeit von der Zeit*, Nachrichten von der Gesellschaft der Wissenschaften zu Göttingen, 1932, p. 97/106; Kunii, S.: *Solutions of cosmological field equations and models of the universe with annihilation of matter*, Memoirs of the College of science, Kyoto Imperial University 15, 1932, p. 97/111; Kohler, M.: *Beiträge zum kosmologischen Problem und zur Lichtausbreitung in Schwerefeldern*, Annalen der Physik 16, 1933, p. 129/161; Robertson, H.P.: *Relativistic cosmology*, Reviews of Modern Physics 5, 1933, p. 62/90; Zaycoff, R.: *Zur relativischen Kosmogonie*, Zeitschrift für Astrophysik 6, 1933, p. 128/137; Mineur, H.: *L'Univers in expansion*, Hermann & C$^{ie}$, Paris 1933; Bronstein, M.P.: *K voprusu o vozmozhnoi teorii mira kak tselogo*, Uspekhi Astronomichskii Nauk 3, 1933, p. 3/30.

[133] The first to mention a "cosmological principle of relativity" was Erwin Finlay Freundlich in discussing a postulate introduced by Edward Milne (see: Freundlich, E.F.: *Ein neuartiger Versuch von E. A. Milne, das kosmologische Problem zu lösen und die Expansion der Spiralnebel zu deuten*, Naturwissenschaften 21, 1933, p. 54/59; Milne adopted then such a terminology in his *Aberystwyth Lectures* of 1933 (see: Milne, E.A.: *Relativity, Gravitation and World Structure*, Clarendon Press, Oxford 1935). In that same year Milne stated the cosmological principle as a general precept according to which: "we refuse to consider our own viewpoint as special, and we regard any other viewpoint as equally good" (Milne, E.A.: *Some points in the Philosophy of Physics*, Philosophy 9, 1934, p. 19/38, on p. 33).

[134] Robertson, H.P.: *Relativistic cosmology...* 1933; Robertson used the expression "Friedmann's universes" in: Robertson, H.P.: *The Expanding Universe*, Science 76, 1932, p. 221/226, on p. 225. In the early 1930's this expression was often reserved for homogeneous and isotropic expanding universes which satisfy the assumption of zero pressure (as in Friedmann's original papers). More often, it was used to define the whole class of homogeneous and isotropic models (with no limitation on the assumptions concerning the matter/energy content or the cosmological constant).

[135] Robertson, H.P.: *Kinematics and World-Structure, part I*, Astrophysical Journal 82, 1935, p. 284/301; *Kinematics and World-Structure, part II*, 83, 1936, p. 187/201; *Kinematics and World Structure, part III*, Astrophysical Journal 83, 1936, p. 257/271. Walker, A.G.: *On Milne's Theory of world-structure*, Proceedings of the London Mathematical Society 42, 1936, p. 90/127.

[136] Merleau Ponty, J.: *Cosmologie du XX siècle*, Gallimard, Paris 1965, p. 74.

Some of them believed it was the task of Mount Wilson's observers[137] to establish which of the ideal universes best fitted the actual one, but others were driven by philosophical and personal beliefs (or prejudices) in limiting the field of possibilities through *a priori* criteria of different kinds. The most clamorous case was probably that of Sir Arthur Eddington who, faithful to the theoretical ideas which spurred him towards a *sui generis* unitary theory of general relativity and quantum mechanics[138], never abandoned the conviction that only Lemaitre's solution of 1927 was indeed appropriate.

The so called Eddington/Lemaitre model describes a closed but ever expanding universe which starts as an Einstein static universe and then expands at an increasing rate as a result of the "cosmical repulsion" due to the presence of a positive cosmological constant. Eddington was an extraordinary popularizer both with his colleagues as well as the general public.

In one of his expositions he vividly pointed out how in the Eddington/Lemaitre universe:

"the radius increases with time in geometrical progression, and ultimately objects must separate at a rate greater than the velocity of light; there is no contradiction in this, for the separation corresponds not to dynamical motion but to inflation of space. Objects separating faster than the velocity of light are cut off from any causal inference on one another, so that in time the universe will become virtually a number of disconnected universes no longer bearing any physical relation to one another"[139].

Similar considerations were then re-presented in the classic "popular" essay entitled *The Expanding Universe*. Here Eddington affirmed:

"I suppose that the distance of one galaxy from the next will ultimately become so great, and the mutual recession so rapid, that neither light nor any other causal influence can pass from one to another. All connection between the galaxies will be broken; each will be a self-contained universe uninfluenced by anything outside it. Such a disintegration is rather a nightmare to conceive; though it does not threaten any particular disaster to human destiny"[140].

In other words: galaxies were destined to become "self contained universes" in an accelerating universe, while today's astronomers had to consider themselves "extraordinarily fortunate" to have arrived "just in time to observe this interesting but evanescent feature of the sky"[141].

---

[137] Hubble will write: "Mathematicians deal with possible worlds, with an infinite number of logically consistent systems. Observers explore the one particular world we inhabit. Between the two stands the theorist. He studies possible worlds but only those which are compatible with information furnished by observers. In other words, theory attempts to segregate the minimum number of possible worlds which must include the actual world we inhabit. Then the observer, with new factual information, attempts to reduce the list further. And so it goes, observation and theory advancing together toward the common goal of science, knowledge of the structure and observation of the universe" (Hubble, E.: *The Problem of the Expanding Universe*, American Scientist 30, p. 99/115, 1942, on p. 105/106).

[138] Cf. Whittaker, E.T.: *Eddington's Principle in the Philosophy of Science,* Cambridge University Press, 1951; Merleau Ponty, J.: *Le Destinée Intellectuelle d'Eddington et sa Signification*, Bullettin de la Société Francaise de Philosophie 54, 1962, p. 1/38; Kilmister, C. W.: *Eddington's search for a fundamental theory. A key to the universe*, Cambridge University Press, Cambridge 1994; Bettini, S.: *Immagini dell'Universo*..., 2001, section II.8 and references quoted therein.

[139] Eddington, A.S.: *The Expansion of the Universe, Report of the Council to the Hundred and Eleventh Annual General Meeting,* Monthly Notices of the Royal astronomical Society 91, 1931, p. 412/416, on p. 415.

[140] Eddington, A. S.: *The Expanding Universe*, Cambridge University Press, Cambridge 1933 (quotation from p. 66/67 of the Penguin edition published in 1940).

[141] Eddington, A.S.: *On the Instability of Einstein's Spherical World*, Monthly Notices of the Royal astronomical Society 90, p. 668/678, on p. 677. This argument was exploited by bishop Barnes against the relativistic interpretation of red shifts and the idea itself of an expanding universe (Barnes, E.W. in: *Contributions to a British Discussion...*1931, p. 721). It was then discussed in: Robertson, H.P.: *Relativistic cosmology...* 1933, p. 72, North, J.D.: *The Measure*...,

Similar conclusions are again fashionable after recent observations of Type Ia supernovae suggesting that the acceleration of the universe is due to a repulsive force analogous to a positive cosmological constant[142]. Ideas connected with these observational data were indeed considered "the hottest topics of discussion among cosmologists during 1998"[143].

Of course this current debate with all its implications stems from a very different source to that of Eddington[144]. Nevertheless many authors continue to use a terminology which resembles that adopted by the Cambridge astronomer in the early 1930's and still talk of how a "universe doomed to accelerate forever will produce a state of growing uniformity and cosmic loneliness"[145].

As we have seen, Eddington used the term inflation to describe the accelerated expansion and referred openly to self-contained universes (moreover, he introduced also the term "bubble" to describe "regions between which no causal influence can ever pass"[146]). In any case, the main motive for his sustaining the Eddington/Lemaitre model lay in theoretical reasons or, better still, in his obstinate belief that the cosmological constant represented a fundamental standard of constancy, indispensable in dealing with lengths in physics.

---

It was fundamentally for his singular ideas that Eddington became one of Henry Dingle's main target when, in a provocative paper appearing in the May 8th 1937 issue of *Nature*, he accused the advocates of a rationalistic approach to cosmology of being "modern Aristotelians" guilty of producing "metaphysics out of mathematics"[147].

Dingle surely embodied a very different tradition in his approach to the science of the universe; a tradition which can include de Sitter amongst his ranks (the Dutch astronomer, in fact, often remarked that the assumptions of homogeneity and isotropy were not "empirical facts" but rather metaphysical hypothesis[148]) and which presumably found its main spokesman in Richard Chase Tolman[149].

Both Tolman and Dingle[150], stated frequently not to be overconfident with the cosmological principle and suggested extreme caution when extrapolating results obtained within the "highly abstract, simplified and idealized" FLRW model[151].

This methodological approach is expressed clearly in Tolman's *Relativity Thermodynamics and Cosmology*, an essay published in 1934 which became an essential textbook for various generations of physicists. He stated that:

"In the first place, we have a natural interest and intellectual pleasure in trying to develop the consequences of any set of mathematical assumptions without reference to possible physical applications. Secondly, since we have based our treatment on acceptable physical theory, we have the right to expect that the theoretical behaviour of our models will at least inform and liberalize our thinking as to conceptual possibilities for the behaviour of the actual universe. In the third place, however, and this is perhaps most important of all, we have the right to hope that the models can be so constructed as to assist in the correlation and explanation of the observed phenomena of the actual universe, and indeed may even be sufficiently representative as to permit some cautious extrapolation forward and backward in time, which will give us not too fallacious ideas as to the past and future history of our surroundings"[152].

During the 1920's and the early 1930's Tolman's main aim consisted in "attacking thermodynamic problems in curved space/time"[153], finally achieving a coherent general relativistic thermodynamics. His investigations reserved important applications to cosmology (for instance, the Caltech physicist explored different kinds of cosmic fluid, revealing the distinctions between a universe dominated by matter and one dominated by radiation[154]).

Tolman's analysis of the problem of the entropy of the universe came then into conflict with classical thermodynamics. He showed that in general relativistic thermodynamics it was possible to avoid the "dreadful final state of quiescence"[155] considering that the increase of entropy is necessarily accompanied by an increase of the proper energy and temperature of each element of a cosmological fluid.

---

[147] Dingle, H.: *Modern Aristotelianism*, Nature 139, 1937, p. 784/786.

[148] Cf. de Sitter, W.: *Some further computations regarding non-static universes*, Bulletin of the Astronomical Institutes of the Netherlands 6, n. 223, 1931, p. 141/145 (in particular p. 144).

[149] Cf. Eisenstaedt, J.: *Cosmology: A Space for Thought on General Relativity*, in: *Foundation of Big Bang Cosmology. Proceedings of the Seminar on the Foundations of Big Bang Cosmology*, Meyerstein, W.F. (ed.), World Scientific, Singapore, 1989, p. 271/295.

[150] Dingle was an associate of Tolman and spent ten months in Pasadena during the Academic Year 1932-1933.

[151] Tolman, R.C.: *Models of the Physical Universe*, Science 75, 1932, p. 367/373, quotation on p. 373.

[152] Tolman, R.C.: *Relativity Thermodynamics and Cosmology*, Clarendon Press, Oxford 1934, on p. 445 of the 1987 Dover edition.

[153] Tolman, R.C.: *On the extension of thermodynamics to general relativity*, Proceedings of the National Academy of Sciences of the U.S.A. 14, 1928, p. 268/272, on p. 268.

[154] Tolman considered a universe filled with radiation just an intellectual exercise (cf.: Tolman, R.C.: *On the theoretical requirements for a periodic behaviour of the universe*, Physical Review 38, 1931, p. 1758/1771, on p. 1767). He never imagined anything like decoupling or any other typical conceptions of the HBB model. In any case, he understood the need to state not only an equation of continuity which describes the relation between pressure and density in a non static universe, but also an equation of state for the fluid. The idea of a universe filled with radiation (in the context of de Sitter's static model) was remarked also in: Silberstein, L.: *Illuminated Spacetime: Optical Effects of Isotropic Radiation Spread over Elliptic Space*, Philosophical Magazine (7) 9, 1930, p. 50/57.

[155] Tolman, R.C.: *Models of the Physical…*, 1932, p. 372.

The ideal tool of Tolman's researches consisted in a model, firstly discussed by Friedmann in 1922[156], "in which the proper volume of the universe increases from zero to an upper limit and returns"[157].

The Russian metereologist adopted a perfect fluid with zero pressure (the so called dust) and treated the cosmological constant $\Lambda$ as a parameter capable of taking any value between $-\infty$ and the positive value $\Lambda_E$ characteristic of Einstein's static universe (so that the larger the value of $\Lambda$, the longer what he called the *Welt Periode* became, tending to infinity for $\Lambda \to \Lambda_E$). For the cases of physical significance Friedmann imagined a cyclical repetition (presupposing then the beginning of a new cycle after zero radius and volume had been reached) which, in an essay of 1923, was compared to the "mythological conceptions of the Hindu"[158].

A proper cyclical model was then presented by the Japanese physicist Tokyo Takeuchi in 1930, who attempted in this way to construct an eternal universe "in agreement with the view of Boltzmann"[159], conjecturing at the same time a minimum radius of finite value rather than zero[160].

But a universe presenting a contracting phase after having reached a maximum expansion gained a certain popularity when Einstein took up cosmology again[161] in May 1931 after years of silence. Although he followed Friedmann's original treatment, filling his model with dust, he nevertheless limited his analysis to the case $\Lambda = 0$.

Contrary to Friedmann, Einstein was extremely worried by the presence of a singularity and hoped that the dismissal of the assumption of homogeneity would have showed a realistic way out of this problem.

The "cycloidal Friedmann/Einstein"[162] universe and some of its variants were then considered by other experts such as Heckmann[163], Lemaitre[164] and de Sitter, with the latter denouncing a "personal idiosyncrasy" towards the "periodically recurring catastrophe" implied by the model[165]. Nevertheless, it was substantially thanks to Tolman if many obscurities about the thermodynamical behaviour of the oscillating model were clarified.

In November 1931 Tolman considered homogeneous universes which expand and contract reversibly at a finite rate without any increase in entropy. He then discarded the Takeuchi model (which involves a negative pressure) as unphysical and proved that contraction to zero proper volume could only be followed by renewed expansion. He, furthermore, showed that a "series of successive expansions and contractions" were inevitable in the Friedmann/Einstein model[166], a conclusion which remained valid even if the universe was filled with radiation (or with any other fluid which does not imply negative pressures). Two months later Tolman concluded that no *thermodynamic hindrance* due to irreversible processes could prevent the succession of cycles of the oscillating model with a null or negative cosmological constant[167]. He showed, moreover, that as a consequence of an increase of entropy, the scale factor must assume a greater maximum value (once maximum expansion has been reached) at any new cycle[168].

---

[156] Friedmann, A.A.: *Über die Krümmung des Raumes*, Zeitschrift für Physik 10, 1922, p. 377/386.

[157] Tolman, R.C.: *On the theoretical requirements…*, 1931, p. 1764.

[158] In any case, Friedmann added that this was just a curiosity Cf. *L'univers comme espace et temps*, in: *Alexandre Friedmann Georges Lemaitre. "Essais de cosmologie"*,…1997, p. 206.

[159] Takeuchi, T.: *On the Cyclic Universe*, Proceedings of the Physico-Mathematical Society of Japan (3) 13, 1931, p. 166/177 (quotation on p. 166).

[160] Indipendently from Tolman, such a possibility was rejected in: Lemaitre, G.: *L'univers en expansion*, Annales de la Société Scientifique de Bruxelles A53, 1933, p. 51/85.

[161] Einstein, A.: *Zum kosmologischen Problem der allgemeinen Relativitätstheorie*, Sitzungberichte der Königlich Preussischen Akademie der Wissenschaften zu Berlin 1931, p. 235/237.

[162] So called in Lemaitre, G.: *L'univers en expansion…*, 1933, p. 83 and, later, in North, J.D.: *The Measure…*, 1965, p. 132 of the Dover edition.

[163] Heckmann, O.H.L.: *Die Ausdehnung der Welt…*, 1932.

[164] Lemaitre, G. *L'expansion de l'espace*, Revue des Questions Scientifiques. (4) 20, 1931, p. 391/410; *L'univers en expansion…*, 1933.

[165] De Sitter remarks were recorded during the *meeting* of the *Royal Astronomical Society* of May 12, 1933 and reported in *Observatory* 56, 1933, p. 173/185 (quotations on p. 184). Cf. also de Sitter, W.: *On the expanding universe and the time scale*, Monthly Notices of the Royal astronomical Society 93, 1933, p. 628/634, on p. 630; *The Astronomical Aspects…, 1933*, p. 185.

[166] Tolman, R.C.: *On the theoretical requirements…*, 1931, p. 1765.

As an appendix of this last paper the American physicist presented, jointly with his student Morgan Ward, a new article[169] to discuss in depth the nature of singularities in the case $\Lambda = 0$. This new contribution appears in retrospect less redundant than his preceding papers, and it is for this reason often considered Tolman's main result[170].

In effect, this paper stated clearly that an exceptional point of null volume was inevitable in the ideal closed universe with no cosmological constant and a physically meaningful fluid (i.e.: stating realistic assumptions on the matter/energy tensor) and that a continued succession of expansions and contractions was then to be expected on physical grounds[171] despite the reversible or irreversible nature of the processes taking place in the cosmic fluid.

The unavoidability of singularities was particularly intriguing for Lemaitre, who invoked the legendary phoenix to describe the behaviour of the oscillating model[172] and, some years later, talked explicitly of a succession of "completely new universes"[173].

Such an image was substantiated in the early 1970's by John Archibald Wheeler who imagined that the beginning of any new cycle of an oscillating universe implied a re-generation of the fundamental constants and of the form of the expansion dynamics, thus depicting the scenario of a temporal sequence of disjointed universes[174].

In any case, until the 1970's, the oscillating closed model with $\Lambda = 0$ was generally considered as the best FLRW universe available both on observational[175] and theoretical grounds. Popular essays were dedicated to it by authors such as Gamow and Öpik[176] and the model was indeed favoured (at least as a "working hypothesis") by Dicke and Peebles in the papers which followed the discovery of cosmic black body radiation[177]. Dicke, as many others before him, was attracted by the oscillating model because of the possibility of avoiding an original creation of matter (or at least relegating it to the infinite past)[178].

---

[167] Tolman, R.C.: *Possibilities in relativistic thermodynamics for irreversible processes without exhaustion of free energy*, Physical Review 39, 1932, p. 320/336.

[168] Cf. Novikov, I.D.-Zel'dovich, Ya. B.: *Physical processes near cosmological singularities*, Annual Review of Astronomy and Astrophysics 11, 1973, p. 387/412; Landsberg, P.T.-Park, D.: *Entropy in an oscillating universe*, Proceedings of the Physical Society of London A346, 1975, p. 485/495; Barrow, J.D.-Dabrowski, M.P.: *Oscillating Universes*, Monthly Notices of the Royal Astronomical Society 275, 1995, p. 850/862.

[169] Tolman, R.C.-Ward, M.: *On the Behaviour of Non-Static Models of the Universe when the Cosmological Constant is Omitted*, Physical Review 39, 1932, p. 835/843.

[170] E.g.: Tipler, F.J.-Clarke, C.-Ellis, G.F.R.: *Singularities and Horizons. A Review Article*, in: *General Relativity and Gravitation*, Held, A. (ed.), Plenum Press, New York 1980, Volume 2, p. 97/206.

[171] In *Relativity Thermodynamics and Cosmology* this result was illustrated advancing a parallel with "the behaviour of an elastic ball, bouncing up and down from the floor" (cf. p. 439 of the Dover edition).

[172] Lemaitre, G.: *L'univers en expansion*…, 1933, p. 85.

[173] See: Godart, O.-Heller, M. (eds.): *The Expanding Universe: Lemaitre's Unknown Manuscript*, Pachart, Tucson 1985, p. 29/30.

[174] Cf. note 20 above.

[175] In the early 1960's astronomers generally considered the value of the mean density in the interval $10^{-31} \leq \rho_0 \leq 10^{-30}$ g cm$^{-3}$. Sandage and collaborators at Mount Palomar then estimated on the basis of quasars counting that $\Omega_0 \approx 2$. Cf. Sandage, A.: *Observational Cosmology*, Observatory 88, 1968, p. 91/106.

[176] Gamow, G.: *Modern Cosmology*, Scientific American 190 (March 1954), p. 53/63; Öpik, E.J.: *Oscillating Universe*. Mentor, New York 1960.

[177] Cf. Dicke, R.H.-Peebles, P.J.E.-Roll, P.G.-Wilkinson, D.T.: *Cosmic Black-Body Radiation*, Astrophysical Journal 142, 1965, p. 414/419 (quotation on p. 415).

[178] Cf. R.H. Dicke in Lightman, A.-Brawer (eds.), R.: *Origins, the lives and worlds of modern cosmologists*, Harvard University Press, Harvard 1990, on p. 205.

Such a hope was indeed frustrated by the singularity theorems of Hawking and Penrose (which demonstrated that singularities were inevitable even if the assumptions of homogeneity and isotropy were discarded, consequently raising doubts as to the physical reality itself of more than one cycle) and later by a series of thermodynamical arguments that seem to forbid the possibility itself of a bounce at the end of the phase of contraction[179].

Finally, both Zeldovich and Novikov and Joseph Silk showed that an infinity of past oscillations was impossible[180] (and that a beginning of time was consequently "unavoidable") on the basis of the finite value of the entropy per baryon number (or specific entropy) in the actual universe[181].

Let us further add that Wheeler was forced to abandon his infinitely cyclical universe (concluding at last that only one cycle was admissible) and that other proposals of a multiverse founded on the basic concept of bouncing closed universes were then advanced by M. A. Markov in the 1980's[182] and, more recently, within studies involving the temporal variation of fundamental[183] constants or within the ambit of cosmological applications of string theory (where Steinhardt and Turok, with their ekpyrotic/cyclic scenarios, claimed to have showed that the universe does not pass through a singularity if we interpret the big crunch/big bang as collisions of branes in the context of M-theory[184]).

Returning to Tolman, he remained, indeed, uncertain about the meaning of the results obtained from the idealized homogeneous model and did not renounce his belief that Einstein's suggestion could in fact be realized[185]. In the paper together with Ward he noted that "the idealization upon which our considerations have been based should be regarded as failing in the neighbourhood of zero volume"[186], while in *Relativity Thermodynamics and Cosmology* he stated explicitly that:

"… the finer details of cosmic behaviour could not in any case be represented by a perfectly homogeneous model. Thus, for example, it should be clearly appreciated that the lower singular state of exactly zero radius, which might be thought of as occurring in the case of an oscillatory time behaviour, must be regarded as the attribute of a certain class of homogeneous models, and not as a state that would necessarily accompany an oscillating expansion and contraction of the whole or parts of the real universe"[187].

---

[179] E.g.: Bludman, S. A. 1984: *Thermodynamics and the end of a closed Universe*, Nature 308, p. 319/322; Guth, A.H.-Sher, M.: *The impossibility of a bouncing universe*, Nature 302, 1983, p. 505/506.

[180] Zel'dovich, YA. B.-Novikov, I.D.: Stroenie i Evolyutsiya Vselennoi [The Structure and Evolution of the Universe], Nauka, Moscow 1975; Silk, J.: *The Big Bang*, Freeman, New York 1989 (in part. p. 391).

[181] Note that in 1976 M. Clutton-Brock advanced a strong anthropic explanation of the actual value of the entropy per baryon number recurring to a universe splitting "into infinitely many branches, or 'worlds', only one of which we can observe". See: Clutton Brock, M.: *Entropy per baryon in 'many worlds' cosmology*, Astrophysics and Space Science 47, 1977, p. 423/433.

[182] On Markov's perpetually oscillating model with "formation of daughter universes" see: Markov, M.A. *Problems of a Perpetually Oscillating Universe*, Annals of Physics 155, 1984, p. 333/57; *Asymptotic freedom and entropy in a perpetually oscillating universe*, Physics Letters 94A, 1983, p. 427/429.

[183] E.g.: Barrow, J.D.-Kimberly, D.-Magueijo, J.: *Bouncing Universes with Varying Constants*, Classical and Quantum Gravity 21, 2004, p. 4289/4296

[184] Khoury, J.– Ovrut, B.-Steinhardt, P.J.-Turok, N.: *Ekpyrotic Universe: Colliding branes and the origin of the hot big bang*, Physical Review D64, 2001, 123522 or ArXiv: hep-th/0103239; Khoury, J.– Ovrut, B.-Steinhardt, P.J.-Turok, N.: *From Big Crunch to Big Bang*, Phys. Rev. D65, 086007, 2002, or:ArXiv: hep-th/0108187; Steinhardt, P.J.-Turok, N.: *A Cyclic Model of the Universe*, ArXiv:hep-th/0111030 (v2 October 2002); Steinhardt, P.J.-Turok, N.: *Cosmic evolution in a cyclic universe*, Physical Review D 65, 2002, 126003or ArXiv: hep-th/0111098; Steinhardt, P.J.-Turok, N.: *The Cyclic Universe: An Informal Introduction*, arXiv:astro-ph/0204479 (April 2002); Turok, N.-Perry, M.-Steinhardt, P.: *M Theory Model of a Big Crunch/Big Bang Transition*, Physical Review D70, 2004, 106004 or: arXiv: hep-th/0408083; Steinhardt, P.J.-Turok, N.: *Beyond Inflation: A Cyclic Universe Scenario*, Physica Scripta T117, 2005, p. 76/85; Steinhardt, P.J.-Turok, N.: *The Cyclic Model Simplified*, New Astronomy Reviews 49, 2005, p.43/57. Note that the proposals quoted in the last two notes represent radical alternatives to inflationary models.

[185] E.g.: Tolman, R.C.: *Relativity Thermodynamics and …*1934, p. 438/439 of the Dover edition.

[186] Tolman, R.C.-Ward, M.: *On the Behaviour…*, 1932, p. 842.

[187] Tolman, R.C.: *Relativity Thermodynamics and …*1934, p. 482 of the Dover edition.

On this basis and convinced of the dangers of "applying to the actual universe any *wide* extrapolations - either spatial or temporal - of results deduced from strictly homogeneous models" Tolman proposed in 1934 a model universe with variable curvature and filled with dust, where "non-interacting zones″ with an open geometry and zones with a closed one (as well as expanding and contracting regions) co-existed[188].

Approximately in the same period Dingle stated that "we have no grounds for supposing that the part of the universe which is observed is typical of the whole"[189] and pointed towards the line element of a "not very inhomogeneous" or "nearly homogeneous" universe of which the FLRW was a special case[190].

In 1936 Dingle adopted the "idea that our expanding system of nebulae is merely a local unit in a larger universe" within the proposal of a universe "majestically quiescent on the grand scale" (i.e. a globally stationary universe where expansion concerns only local regions or particular epochs)[191].

Conceptions of a universe which is inhomogeneous on a large scale, apart from the fundamental contributions of Lemaitre[192], were typical of Soviet authors (from Moris Semenovich Eigenson, who described in the 1930's an infinite universe where the expansion was a local occurrence, to Shirokov and Fisher[193]) who shared with Tolman[194] the need of avoiding a special beginning (or simply an origin) of the whole world.

With regard to this last subject, an interesting proposal of a "self-perpetuating" inhomogeneous universe with density fluctuations on all the observable scales was presented by Ronald Gordon Giovanelli in 1963[195].

The Australian physicist imagined that the average density and expansion rate of the observed universe "need not be representative of the universe as a whole" and aimed then to find compatibility, on a statistical basis, between inhomogeneity and the "aesthetic attractiveness" of steady-state cosmology. He pictured an eternal, infinite universe where "the time-averaged properties of any one region may be the same for all parts of the universe, though at any one time the properties of individual regions might differ greatly".

Without quoting Tolman and Dingle, Giovanelli similarly presented a picture where "some regions may be expanding, others contracting" and where " this state of affairs may reverse from time to time". He asked, moreover, in what way our observable region of the universe could be defined "atypical" suggesting that

"a description of the universe beyond our "observable" region is not merely of philosophical interest but is of direct importance for understanding the dynamics of, and the density distribution within, our observable region"[196].

---

[188] Tolman, R.C.: *Effects of Inhomogeneity on Cosmological Models*, Proceedings of the National Academy of Sciences of the U.S.A. 20, 1934, p. 169/176. Quotations from p. 176 and p. 175.

[189] Dingle, H.: *On isotropic models of the universe with special reference to the stability of the homogeneous and static states*, Monthly Notices of the Royal Astronomical Society 94, 1933, p. 134/158, p. 157.

[190] Ibid. p. 142. This line element was introduced in: Dingle, H.: *Values of $T_m{}^n$ and the Christoffel Symbols for a Line-Element of Considerable Generality*, Proceedings of the National Academy of Sciences of the U.S.A. 19, p. 559/563.

[191] Dingle, H.: *Physical Universe*, Journal of the Washington Academy of Sciences 26, 1936, p. 183/195 (quotations on pages 193 and 194).

[192] In particular Lemaitre, G.:*La formation des nébuleuses dans l'univers in expansion* Compte Rendu Hebdomadaire des Scéances del'Académie des Sciences 196, 1933, p. 1085/1087. On the history of inhomogeneous models, cf. Krasinski, A.: *Inhomogeneous Cosmological Models*, Cambridge University Press, Cambridge 1997.

[193] Eigenson, M.S.: *Bol'shaia Vselennaia: ochrk sovremennykh znanii o vnegalakticheskikh tumannostiakh*, Izd-vo Akademii nauk, Moscow 1936; Shirokov, M.F.-Fisher, I.Z.: *Isotropic space with discrete gravitational-field. On the theory of a nonhomogeneous isotropic universe*, Soviet Astronomy-AJ 6, 1963, p. 699/705.

[194] Cf. Tolman, R.C.: *Relativity Thermodynamics and* …1934, p. 485 of the Dover edition; *The age of the universe from the red shift in the spectra of extragalactic objects*, Proceedings of the National Academy of Sciences of the U.S.A. 47, 1935, p. 202/203; on p. 202; *The Age of the Universe*, Reviews of Modern Physics 21, p. 374/378.

[195] Giovanelli, R. G.: *A Fluctuation Theory of Cosmology*, Monthly Notices of the Royal Astronomical Society 127, 1964, p. 461/469. On Giovanelli see: http://www.austehc.unimelb.edu.au/guides/giov/giovanelli.htm .

In recent times such a question has assumed considerable relevance after the affirmation that the large structure of the universe could derive from quantum fluctuations occurring during the inflationary phase and, since then, redshifted well beyond the particle horizon.

Such an idea has also been exploited to point towards a mechanism (i.e.: a backreaction effect of super-horizon perturbations) capable of explaining the presumed acceleration of the universe as an alternative to dark energy[197].

## 5.JUST A BRIEF CONCLUSION

Andrei Linde has recently confessed his belief in inflationary theory since it is "by now … 20 years old" while the "typical lifetime of a new trend in high energy physics and cosmology nowadays [is] 5 to 10 years"[198]. His optimism is, in my mind, hard to share, although legitimised all the same since mathematically founded speculations must be considered essential to physical progress.

In any case, what I find remarkable in current debates on the multiverse is the persistence of old issues such as those concerning the relation between laws and initial conditions (or those related to what is necessary and what is accidental, which is by and large the same) or the controversies concerning antinomies such as finite/infinite, originated in time/eternal, globally evolving/globally stationary.

These topics seem to recur again and again and each time they present themselves on increasingly deeper levels of physical knowledge. They were of principal relevance to the HBB/steady-state controversy and were then shifted to the very outset of the evolutionary history of the universe, to then pass finally onto the multiverse scenarios.

One could simply conclude that since the "Universe" is a name given to the most inclusive and comprehensive physical system, talk of many universes is surely purely non-sense[199]. However, as Lewis Feuer noted in 1933, if one assumes that "there is nothing necessary about a physical universe", it follows that "speculations about the existence of universes with different laws of physics are legitimate"[200].

ACKNOWLEDGEMENTS

---

I am very grateful to Pierre Kerszberg for his appraisal of my work I am also greatly indebted to Alberto Cappi, Svitlana Hluvko, Milan Cirkovic for their helpful suggestions: At last, I will always be grateful to Paolo Rossi and Silvio Bergia.